\documentclass[twocolumn]{aastex701}

\usepackage{newtxtext,newtxmath}
\usepackage[T1]{fontenc}
\usepackage{graphicx}
\usepackage{amsmath,amssymb,mathtools}
\usepackage{bm}
\usepackage{booktabs}
\usepackage{multirow}
\usepackage{placeins}
\usepackage{hyperref}
\usepackage{xcolor}

\newcommand{\ud}{\mathrm{d}}
\newcommand{\rg}{r_{\rm g}}
\newcommand{\tg}{t_{\rm g}}
\newcommand{\slow}{r_{\rm slow}}
\newcommand{\base}{t_{\rm base}}
\newcommand{\Mc}[1]{\mathcal{#1}}
\newcommand{\tblcell}[2]{\parbox[t]{#1}{#2}}

\begin{document}

\title{CoportSL: A Contribution-constrained Hybrid Slow-light Framework for Time-dependent Polarized GRMHD Imaging}
\shorttitle{CoportSL Hybrid Slow-light Imaging}
\shortauthors{Zhou et al.}

\correspondingauthor{Minyong Guo}
\email{minyongguo@bnu.edu.cn}

\correspondingauthor{Bin Chen}
\email{chenbin1@nbu.edu.cn}

\author[0009-0001-0796-1547]{Fan Zhou}
\email{202631101012@mail.bnu.edu.cn}
\affiliation{School of Physics and Astronomy, Beijing Normal University, Beijing 100875, China}

\author[0009-0002-2360-2971]{Jiewei Huang}
\email{jieweihuang@mail.bnu.edu.cn}
\affiliation{School of Physics, Peking University, No. 5 Yiheyuan Road, Beijing 100871, China}

\author[0009-0007-4339-0570]{Yuehang Li}
\email{adamdarx@qq.com}
\affiliation{School of Physics, Peking University, No. 5 Yiheyuan Road, Beijing 100871, China}

\author[0000-0001-5577-575X]{Minyong Guo}
\email{minyongguo@bnu.edu.cn}
\affiliation{School of Physics and Astronomy, Beijing Normal University, Beijing 100875, China}
\affiliation{Key Laboratory of Multiscale Spin Physics (Beijing Normal University), Ministry of Education, Beijing 100875, China}

\author[0000-0003-4509-9705]{Bin Chen}
\email{chenbin1@nbu.edu.cn}
\affiliation{Institute of Fundamental Physics and Quantum Technology, \& School of Physical Science and Technology, Ningbo University, Ningbo, Zhejiang 315211, China}
\affiliation{School of Physics, Peking University, No. 5 Yiheyuan Road, Beijing 100871, China}

\begin{abstract}
Fast-light approximations neglect fluid evolution along rays, whereas slow-light modeling is indispensable for recovering the true magnetohydrodynamic state. However, full slow-light radiative transfer for extended general relativistic magnetohydrodynamic (GRMHD) sources requires simultaneous access to many fluid snapshots and is memory-intensive. We introduce CoportSL, the first contribution-constrained hybrid slow-light framework for time-dependent full-Stokes imaging. It uses emission, absorption, and Faraday contributions to identify where fluid evolution must be retained, applies fast light elsewhere, and loads only snapshots spanning the relevant delays. Tests with M87*-like magnetically arrested disk GRMHD data show that the contribution-based region and delay-based snapshot restrictions each keep normalized full-image Stokes differences below $4\times10^{-3}$ relative to the corresponding complete calculation. At this accuracy, CoportSL requires $75.3\%$ and $44.7\%$ fewer snapshot layers for near-horizon and jet images, respectively; its per-frame slow-light transfer time remains comparable to fast light. For the two configurations, source-code estimates place the capacities of the principal data structures at $255$--$657\,\mathrm{GiB}$ for a fixed public ipole version and $20.2$--$37.3\,\mathrm{GiB}$ for CoportSL, bringing both configurations within workstation-scale memory. Fast--slow comparisons further show close agreement in near-horizon variability, whereas jet variability follows similar overall trends but differs in local peaks and amplitudes; in both cases, fast light misses substantial full-Stokes spatial structure. As the next-generation Event Horizon Telescope (ngEHT) advances toward dynamical imaging and spatially resolved polarimetry, CoportSL provides a computationally practical way to model full-Stokes finite-light-travel-time signatures in extended black hole systems.
\end{abstract}


\section{Introduction}\label{sec:intro}

The Event Horizon Telescope (EHT) has obtained horizon-scale total-intensity and polarization images of M87* and Sgr A* and has systematically analyzed their physical origins, black hole shadows, magnetic-field structures, and polarization properties \citep{M87_1,M87_4,M87_5,M87_7,M87_8,M87_9,SgrA_1,SgrA_3,SgrA_4,SgrA_5,SgrA_7,SgrA_8,M87_2018_I,M87_2018_II}. Using these observations to constrain the strong-gravity spacetime and magnetized accretion flow near a black hole requires a theoretical chain from fluid evolution to observables. General relativistic magnetohydrodynamics (GRMHD) provides the dynamical evolution of magnetized accretion flows in the strong-gravity regime \citep{EventHorizonTelescope:2019pcy,Gammie_2003}. Coupling GRMHD to general relativistic radiative transfer (GRRT) produces synthetic millimeter images and polarization observables from numerical accretion flows and electron thermodynamics \citep{Broderick:2003fc,Shcherbakov:2010kh,Dexter:2016cdk,Moscibrodzka:2017lcu,Pu:2018ute,Younsi:2019iee,Bronzwaer:2020kle,Aimar:2023vcs,noble2007simulating}. These theoretical predictions can be compared directly with EHT observations to constrain the accretion state, viewing geometry, accretion rate, and electron-thermodynamic parameters \citep{M87_5,M87_8,SgrA_5,SgrA_8,M87_2018_II}, and to test departures of the black hole spacetime from Kerr predictions through the black hole shadow and near-horizon polarization structure \citep{EventHorizonTelescope:2020qrl,EventHorizonTelescope:2021dqv,SgrA_6,Hou:2024qqo}.

Horizon-scale black hole imaging is expanding from time-averaged images to time-dependent images and polarized light curves \citep{Conroy:2023,Najafi-Ziyazi:2023oil,EventHorizonTelescope:2025vum,Zhou_2026}. Multifrequency time lags and photon-ring-related statistics are also emerging as important observables connecting time-dependent structures to propagation effects \citep{Vos:2024,Jiang:2025huk,Hadar:2021,Wong:2024Echoes,Zhang:2025vyx,Bezdekova:2026}. This development requires theoretical models not only to reproduce average images but also to describe correctly the temporal relation between successive images. Finite light-travel time must therefore be accounted for: radiation received at one observer time may originate from different locations and different source times, and propagation delays can alter both image structure and its evolution \citep{Bronzwaer:2018,RojasPaternina:2026}. For example, the fast-light approximation introduces a systematic bias in the pattern speed of GRMHD movies of Sgr A* \citep{Conroy:2023}; in studies of photon rings and light echoes, the relative delays between image orders are themselves the target signals \citep{Hadar:2021,Wong:2024Echoes,Zhang:2025vyx,Bezdekova:2026}.

Many existing GRMHD image libraries use the fast-light approximation \citep{moscibrodzka2009radiative,Gold:2016hld}. In this approximation, all samples in a synthetic image read the same GRMHD snapshot, although the photon trajectories are still computed in curved spacetime. When the emission region is compact and the fluid-evolution timescale exceeds the propagation time along the relevant rays, fast light generally provides reliable results while substantially reducing data access for long image sequences \citep{Bronzwaer:2018,Moscibrodzka:2021}. Its applicability must be reassessed, however, if the fluid evolves appreciably during photon propagation or if the target observable depends directly on relative delays among different rays or image orders. In contrast, slow-light GRRT retains the coordinate time along each photon path and reads the GRMHD fluid state at the corresponding source time at every sample, so a single image generally requires simultaneous access to multiple snapshots \citep{Bronzwaer:2018,White:2022}. For an extended three-dimensional source, the required time span grows with the propagation-delay range, increasing both snapshot access and memory consumption \citep{Vos:2024,RojasPaternina:2026}.

Previous work has followed three main approaches to slow-light calculations. The first targets compact near-horizon source regions and reads fluid snapshots at the appropriate source times along the complete radiative-transfer path. Because the propagation-delay range within the source is limited, the number of snapshots required for one image remains manageable; early post-processing of Sgr A* and time-dependent GRRT with RAPTOR follow this approach \citep{Dexter2010,Bronzwaer:2018}. The second retains the full computational domain and its complete propagation-delay range, reconstructing the fluid state from the nearest snapshot or by linear interpolation between adjacent snapshots. Blacklight and ipole-based full-Stokes fast- and slow-light calculations provide implementations of this approach \citep{White:2022,Moscibrodzka:2021}. The third reduces the data range required by full slow light through prescribed spatial boundaries or delay intervals. Its spatial implementation is commonly termed hybrid slow light: a slow-light region is carved out of the complete source, fluid evolution is retained only within that region, and fast light is used outside it. Such methods cannot, however, determine the extent of the slow-light region in advance. In practice, the region is usually expanded from a small initial domain and the imaging repeated until further expansion no longer appreciably changes the target observable, thereby establishing convergence. \citet{Vos:2024} divided a GRMHD source using a hybrid fast/slow-light radius and calculated multifrequency time lags with a finite number of snapshots. The temporal implementation instead shortens the source-time range covered by each frame; for example, \citet{RojasPaternina:2026} used the dominant propagation-delay interval for semianalytic black hole movies.

For an extended polarized GRMHD source, determining the slow-light region through repeated domain expansion makes the calculation cumbersome. Moreover, existing hybrid schemes primarily target total intensity or particular time-lag observables, and selecting the slow-light region solely from the geometric extent of the emitter does not cover all regions important to polarized propagation. A region with weak direct emission can still alter polarized radiation arriving from elsewhere through absorption, Faraday rotation, and Faraday conversion \citep{Shcherbakov:2010kh,Dexter:2016cdk,Moscibrodzka:2017lcu}. A hybrid slow-light method for an extended polarized source must therefore answer two coupled questions: which spatial regions must retain fluid evolution, and how much GRMHD time is required by those regions?

To address these questions, we present CoportSL, a slow-light extension of Coport-2.0 \citep{Huang2024,Zhou_2026} designed for a fixed background spacetime and spatial grid. Rather than repeatedly expanding the slow-light region and rerunning the imaging calculation, CoportSL first uses a low-cost pre-analysis to evaluate the contributions of different spatial regions to full-Stokes radiative transfer and automatically constructs the slow-light region for prescribed contribution tolerances. It then determines an effective time window from the propagation delays within that region. The region boundary is thus obtained directly from the spatial support of the transfer coefficients, without using multiple production slow-light images as a selection step.

This paper is organized as follows. Section~\ref{sec:method} describes the CoportSL slow-light radiative-transfer framework, including polarized transfer, GRMHD time sampling, slow-light time mapping and interpolation, the spatial region and effective window, and the complete computational procedure. Section~\ref{sec:setup} uses M87*-like GRMHD data to determine the time sampling, select the region and window, and benchmark the accuracy and performance. Section~\ref{sec:results} compares fast- and slow-light observables, and Section~\ref{sec:summary} presents the summary and discussion. Lengths and times are expressed in units of $\rg=GM/c^2$ and $\tg=GM/c^3$, respectively.

\section{Slow-light Radiative-transfer Framework}\label{sec:method}

\subsection{Radiative-transfer Equation}\label{subsec:rt}

Each image-plane pixel corresponds to a null geodesic traced backward from the observer. Let $x^\mu$ denote the spacetime coordinates of a photon event, $\lambda$ the affine parameter, $k^\mu=\ud x^\mu/\ud\lambda$ the photon wave vector, and $\Gamma^\mu_{\alpha\beta}$ the Christoffel symbols of the background metric, with Greek indices running from 0 to 3. The geodesic, wave vector, and polarization reference vector $f^\mu$ evolve according to \citep{Huang2024,Bronzwaer:2020kle}
\begin{equation}
\begin{aligned}
\frac{\ud x^\mu}{\ud\lambda}&=k^\mu,\\
\frac{\ud k^\mu}{\ud\lambda}
&=-\Gamma^\mu_{\alpha\beta}k^\alpha k^\beta,\\
\frac{\ud f^\mu}{\ud\lambda}
&=-\Gamma^\mu_{\alpha\beta}k^\alpha f^\beta .
\end{aligned}
\end{equation}
The polarization reference vector $f^\mu$ is orthogonal to $k^\mu$ and is parallel transported along the same geodesic.

Let $u^\mu$ be the fluid four-velocity, $\nu=-k_\mu u^\mu$ the photon frequency measured in the fluid comoving frame, and $I_\nu$, $Q_\nu$, $U_\nu$, and $V_\nu$ the Stokes specific intensities at that frequency. CoportSL evolves the Lorentz-invariant Stokes vector $\vec{\Mc S}=(\Mc I,\Mc Q,\Mc U,\Mc V)^{\rm T}$, where $\Mc I=I_\nu/\nu^3$, $\Mc Q=Q_\nu/\nu^3$, $\Mc U=U_\nu/\nu^3$, and $\Mc V=V_\nu/\nu^3$. The parallel-transported $f^\mu$ defines a polarization basis propagated along the geodesic. Let $b^\mu$ denote the comoving magnetic-field four-vector, and define the photon propagation direction in the comoving frame and the two-dimensional screen projector orthogonal to $u^\mu$ and $k^\mu$ by \citep{Huang2024, Zhou_2026}
\begin{equation}
e_k^\mu=\frac{k^\mu}{\nu}-u^\mu,
\qquad
P^{\mu\nu}=g^{\mu\nu}+u^\mu u^\nu-e_k^\mu e_k^\nu.
\end{equation}
Projecting $f^\mu$ and $b^\mu$ onto this screen gives the signed angle between the parallel-transported polarization basis and the local magnetic-field basis: \citep{Huang2024, Zhou_2026}
\begin{equation}
\begin{aligned}
\chi={}&\operatorname{sgn}\!\left(\varepsilon_{\mu\nu\rho\sigma}
u^\mu f^\nu b^\rho k^\sigma\right)\\
&\times\arccos\!\left[
\frac{P^{\mu\nu}f_\mu b_\nu}
{\sqrt{\left(P^{\mu\nu}f_\mu f_\nu\right)
\left(P^{\alpha\beta}b_\alpha b_\beta\right)}}
\right].
\end{aligned}
\label{eq:chi}
\end{equation}
Here $\varepsilon_{\mu\nu\rho\sigma}$ is the Levi--Civita tensor, and the sign of the oriented four-volume distinguishes the two possible rotation orientations between the bases. Because a linear-polarization basis is unchanged by a rotation through $\pi$, $\chi$ is understood modulo $\pi$. The corresponding Stokes rotation matrix is \citep{Huang2024,Bronzwaer:2020kle}
\begin{equation}
R(\chi)=
\begin{pmatrix}
1&0&0&0\\
0&\cos 2\chi&-\sin 2\chi&0\\
0&\sin 2\chi&\cos 2\chi&0\\
0&0&0&1
\end{pmatrix}.
\end{equation}
In the local magnetic-field basis, let $j_{\nu,A}$ and $\alpha_{\nu,A}$ ($A\in\{I,Q,U,V\}$) denote the emission and absorption coefficients at the comoving frequency $\nu$, and let $\rho_{\nu,A}$ ($A\in\{Q,U,V\}$) denote the Faraday coefficients at the same frequency. The corresponding invariant coefficients are $j_A=j_{\nu,A}/\nu^2$, $\alpha_A=\nu\alpha_{\nu,A}$, and $\rho_A=\nu\rho_{\nu,A}$. For the gyrotropic electron distributions adopted here, whose momentum distributions are axisymmetric about the local magnetic-field direction, the emission vector and propagation matrix in this basis are \citep{Shcherbakov:2010kh,Dexter:2016cdk}
\begin{equation}
\vec j=
\begin{pmatrix}
j_I\\ j_Q\\ 0\\ j_V
\end{pmatrix},
\qquad
\mathbf K=
\begin{pmatrix}
\alpha_I & \alpha_Q & 0 & \alpha_V\\
\alpha_Q & \alpha_I & \rho_V & 0\\
0 & -\rho_V & \alpha_I & \rho_Q\\
\alpha_V & 0 & -\rho_Q & \alpha_I
\end{pmatrix}.
\end{equation}
Here $j_Q$ and $\alpha_Q$ describe linearly polarized emission and absorption in the local magnetic-field basis, $j_V$ and $\alpha_V$ describe circularly polarized emission and absorption, $\rho_V$ describes Faraday rotation, and $\rho_Q$ describes Faraday conversion. The choice of basis gives $j_U=\alpha_U=\rho_U=0$. The full-Stokes radiative-transfer equation is \citep{Broderick:2003fc,Shcherbakov:2010kh,Dexter:2016cdk}
\begin{equation}
\frac{\ud\vec{\Mc S}}{\ud\lambda}
=R(\chi)\vec j-
\left[R(\chi)\mathbf K R(-\chi)\right]\vec{\Mc S}.
\label{eq:transfer}
\end{equation}
Within each discrete ray segment, CoportSL treats the local transfer coefficients at the sample as constant and advances the Stokes vector to the next sample using the analytic solution of the constant-coefficient full-Stokes transfer equation \citep{degl1985solution}. After integration over the complete path, Stokes $Q$ and $U$ in the parallel-transported basis are rotated into the fixed polarization reference basis on the observer's image plane, yielding a full-Stokes image. The local coefficients depend on the density, internal energy, fluid velocity, magnetic field, and electron distribution and must therefore be evaluated from the GRMHD fluid state at each ray sample. A slow-light calculation additionally requires their evolution in coordinate time, so the GRMHD data must provide both spatial localization and temporal sampling.

\subsection{GRMHD Data Preparation and Preprocessing}\label{subsec:snapshot-cadence}

We consider GRMHD data ordered by a common coordinate time and sharing the same stationary background spacetime and fixed spatial grid. The fixed grid may be uniform or may use static mesh refinement (SMR), in which the grid hierarchy does not change with time \citep{Porth2017}. For an imaging setup with a fixed observer, camera, and background geometry, the geodesics and their polarization bases can be reused throughout the image sequence. The fixed grid further allows reuse of the grid locations and spatial-interpolation weights of the ray samples.

CoportSL first integrates each geodesic backward from the observer and records the discrete path required for the subsequent forward solution of Equation~\eqref{eq:transfer}. The geodesic step size is constrained by both the adaptive-integration error and the local grid scale, such that a ray takes at least two integration steps while crossing one local grid-cell scale. Each sample $i$ stores $x_i^\mu$, $k_i^\mu$, $f_i^\mu$, the affine step $\Delta\lambda_i$, the grid and cell indices, and the spatial-interpolation weights. The forward radiative-transfer calculation reuses this information directly.

The preprocessing above supplies the geometric and spatial-sampling information of the rays, but slow light also requires a time-dependent fluid state at each sample. We use GRMHD snapshots at strictly increasing coordinate times $t_n$ with a fixed interval between adjacent outputs and denote the $n$th snapshot by $\mathbf P(\mathbf x,t_n)$. The primitive-variable vector $\mathbf P$ contains the density, internal energy, spatial velocity, magnetic field, and other quantities needed to reconstruct the local fluid state. Because slow light requires simultaneous access to multiple snapshots, data preparation must balance accuracy against memory and storage costs. This balance concerns both the storage precision of the primitive variables and ray information and a snapshot interval capable of resolving changes in the fluid and observables.

Storage precision must balance the reliability of the simulation evolution against the memory cost of slow light. GRMHD solves highly nonlinear evolution equations, for which small roundoff errors accumulate during successive time steps and gradually separate the later fluid states. The simulation evolution and original outputs therefore use double precision. GRRT instead post-processes existing snapshots, and its accuracy is limited primarily by data discretization and subsequent spatial and temporal interpolation. In the present calculation, these errors are much larger than the roundoff introduced by single-precision storage. CoportSL therefore stores ray-sample information, spatial-interpolation weights, and primitive variables read from the snapshots in single precision, allowing more of the snapshots required by slow light to reside in the same amount of memory.

The snapshot interval must balance the accuracy of fluid reconstruction against data costs. A fast-light calculation makes every sample in one image read the same fluid time. Slow light retains photon propagation time, so different samples generally require different fluid times and $\mathbf P$ must be reconstructed between adjacent snapshots \citep{Bronzwaer:2018,White:2022}. An excessively large snapshot interval increases scalar-interpolation errors; when a vector changes direction between adjacent snapshots, componentwise linear interpolation also reduces the magnitude of the reconstructed vector. An interval that is too small, however, rapidly increases snapshot storage, data access, and the memory required for simultaneously loaded snapshots.

Candidate snapshot intervals are tested at both the fluid and observable levels. At the fluid level, we retain snapshots from a short, high-cadence GRMHD sequence at several candidate intervals, reconstruct the primitive variables at the omitted times using the same linear scheme as in the production slow-light calculation, and compare the reconstruction with the original sequence. In addition to scalars such as density and internal energy, vector magnitudes such as the magnetic-field strength require checks of both the normalized absolute difference and the signed bias that indicates whether the reconstruction is systematically high or low. At the observable level, we use the same high-cadence sequence to generate fast-light images, retain images at the same candidate intervals, and linearly interpolate each Stokes image pixel by pixel between adjacent retained times. Differences in the interpolated images and their total fluxes relative to the original sequence determine whether a candidate interval captures the observable variability. The fluid- and image-level tests jointly determine the production snapshot interval.

Once the production snapshot interval has been determined, a long GRMHD snapshot sequence can be prepared at that cadence. The emission, absorption, and Faraday coefficients must be evaluated from local physical quantities in Gaussian units, so the primitive variables stored in code units must be converted to Gaussian units and an electron distribution must be specified. The model flux obtained after this conversion need not match the target observation. The scale invariance of the GRMHD equations, however, allows the snapshot primitive variables to be multiplied by a scale factor without changing the fluid evolution. We refer to adjusting this factor until the model flux matches the target value as calibration. Specifically, fast-light images are first generated over an interval with relatively steady fluid conditions and their mean flux is evaluated; the scale factor is then iterated until the model mean matches the target flux. CoportSL provides accompanying automatic calibration code for the fast-light imaging, mean-flux evaluation, and scale-factor iteration over the steady interval. The resulting scale factor and electron distribution are used in both the regional-contribution analysis and the production fast- and slow-light calculations.

\subsection{Slow-light Time Mapping and Interpolation}\label{subsec:slow-scheme}

After the snapshot sequence and radiative setup have been prepared, a slow-light calculation must determine the fluid time associated with each sample in an image and identify the two snapshots that bracket that time. At a fixed observer time, a slow-light image contains radiation from different source times, as set by the coordinate propagation delays along the null geodesics \citep{Bronzwaer:2018,RojasPaternina:2026}. To map relative propagation delays to GRMHD coordinate time, CoportSL labels each image with an available snapshot time $\base$ and selects a reference region $\Omega_{\rm ref}$ that remains fixed throughout the image sequence and contains ray samples to define the delay zero point. Let $x_{\rm obs}^0$ and $x_i^0$ be the time coordinates of the observer event and sample event, respectively, and define the coordinate propagation delay from the sample to the observer as $d_i=x_{\rm obs}^0-x_i^0$. The sample with the minimum $d_i$ in the reference region defines the zero point of the propagation delay and reads $\base$; all other samples are shifted toward earlier or later times relative to this anchor. We define
\begin{equation}
d_{\rm ref}=\min_{i\in\Omega_{\rm ref}}d_i,
\qquad
\Delta t_i=d_{\rm ref}-d_i .
\label{eq:time-offset}
\end{equation}
For an image with numerical label $\base$, the ideal slow-light fluid time is
\begin{equation}
t_{{\rm fluid},i}=\base+\Delta t_i .
\label{eq:fluid-time}
\end{equation}
Samples with $\Delta t_i<0$ require snapshots before $\base$, whereas samples with $\Delta t_i>0$ require snapshots at later times. This convention may be understood by first anchoring the sample with the smallest propagation delay in the reference region to one GRMHD snapshot, and then using the propagation delay of every other sample relative to that anchor to determine its required fluid time. A given image sequence uses a fixed $\Omega_{\rm ref}$ and a fixed labeling convention for $\base$.

Given $t_{{\rm fluid},i}$, we locate adjacent times $t_a$ and $t_b$ in the ordered snapshots such that $t_a\leq t_{{\rm fluid},i}\leq t_b$. After spatial interpolation within each snapshot, the primitive variables at sample $i$, $\mathbf P_i$, are reconstructed using the linear scheme commonly adopted in existing slow-light calculations \citep{Bronzwaer:2018,White:2022}:
\begin{equation}
\mathbf P_i=(1-\eta_i)\mathbf P(\mathbf x_i,t_a)+\eta_i\mathbf P(\mathbf x_i,t_b),
\qquad
\eta_i=\frac{t_{{\rm fluid},i}-t_a}{t_b-t_a}.
\label{eq:interp}
\end{equation}
Here $\eta_i$ is the temporal-interpolation weight for sample $i$. The magnetic field enters $\mathbf P$ through its components. When its direction changes between adjacent snapshots, the linear combination of these components systematically reduces the reconstructed field magnitude relative to the true evolution. Local quantities such as density, temperature, comoving magnetic field, and comoving-frame photon frequency are then calculated from the interpolated state, so the error propagates into the nonlinear emission, absorption, and Faraday coefficients \citep{Dexter2010,White:2022}. The production snapshot cadence must constrain both the primitive-variable interpolation error and the resulting observable error.

The radiative coefficients are evaluated after the spatial and temporal interpolation of $\mathbf P_i$, and Equation~\eqref{eq:transfer} is then advanced along the complete discrete path. The base time $\base$ organizes snapshot access and labels the output image; the common time reference of the fast- and slow-light sequences is determined separately when the two sequences are compared.

\begin{figure*}[t]
\centering
\includegraphics[width=0.98\textwidth]{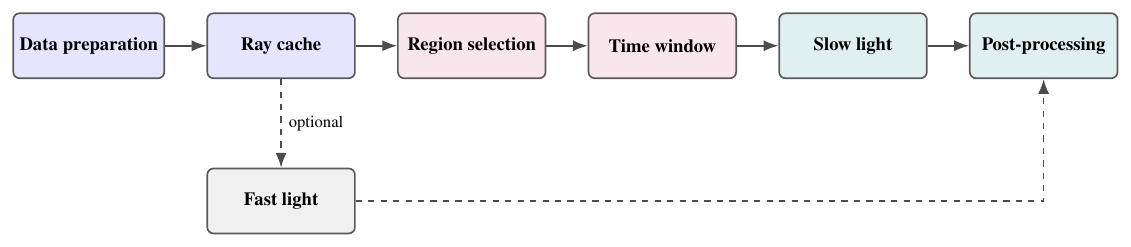}
\caption{Computational procedure of CoportSL. Blue nodes denote data preparation and ray caching, purple nodes denote the selection of the slow-light region and time window, and cyan nodes denote slow-light radiative transfer and observable post-processing. The gray dashed branch is an optional fast-light calculation that can be compared with the slow-light sequence during post-processing.}
\label{fig:workflow}
\end{figure*}

\subsection{Hybrid Slow-light Scheme and Effective Time Window}\label{subsec:region-general}

Full slow light requires a single image to access the GRMHD time range covered by all ray samples \citep{Bronzwaer:2018,RojasPaternina:2026}. Hybrid slow light first identifies the spatial regions containing the dominant radiative-transfer coefficient contributions and then evaluates the propagation delays within those regions, thereby reducing the number of three-dimensional primitive-variable snapshots required simultaneously by one image.

We first partition the complete source domain $\Omega_{\rm source}$ covered by the radiative-transfer integration into a set of disjoint elementary regions,
\begin{equation}
\Omega_{\rm source}=\bigcup_k\Mc R_k,
\qquad
\Mc R_k\cap\Mc R_l=\varnothing\quad(k\ne l),
\end{equation}
where $\Mc R_k$ denotes the $k$th elementary region, whose shape may be defined according to the source structure and scientific objective. The region selection uses six scalar radiative-transfer coefficients. For the gyrotropic electron distributions adopted in Section~\ref{subsec:rt}, $j_U=\alpha_U=\rho_U=0$ in the local magnetic-field basis, and we define
\begin{equation}
\begin{aligned}
j_P&=\left(j_Q^2+j_U^2+j_V^2\right)^{1/2}
=\left(j_Q^2+j_V^2\right)^{1/2},\\
\alpha_P&=\left(\alpha_Q^2+\alpha_U^2+\alpha_V^2\right)^{1/2}
=\left(\alpha_Q^2+\alpha_V^2\right)^{1/2},\\
\rho_C&=\left(\rho_Q^2+\rho_U^2\right)^{1/2}=|\rho_Q|.
\end{aligned}
\label{eq:coeff-amplitudes}
\end{equation}
which characterize the amplitudes of total polarized emission, polarized absorption, and Faraday conversion, respectively; Faraday rotation is characterized by $\rho_V$. Because $R(\chi)$ is an orthogonal rotation in the Stokes $Q$--$U$ plane, these amplitudes are identical in the local magnetic-field and parallel-transported polarization bases. We denote the coefficient set used for region selection by
\begin{equation}
\mathcal X=\{j_I,j_P,\alpha_I,\alpha_P,\rho_V,\rho_C\}.
\end{equation}
For any $X\in\mathcal X$, the path-weighted contribution of region $\Mc R_k$ at time $t$ is defined as
\begin{equation}
C_X(\Mc R_k;t)
=
\frac{\sum_{i\in\Mc R_k}|X_i(t)|w_i}
{\sum_{i\in\Omega_{\rm source}}|X_i(t)|w_i},
\label{eq:region-support}
\end{equation}
where $X_i(t)$ is the local coefficient at sample $i$ and time $t$, $p(i)$ is the image-plane pixel containing that sample, and the path weight is $w_i=\omega_{p(i)}\Delta\lambda_i$. Here $\omega_{p(i)}$ is the pixel weight in the image integral. We use a regular image grid with equal pixel weights, so $w_i=\Delta\lambda_i$. The absolute value makes $C_X$ measure the fractional support of the local coefficient amplitude in each region; coefficients with nonzero denominators participate in the region selection. The quantity $C_X$ is a pre-analysis statistic for the spatial selection. Because emission, absorption, and Faraday effects are coupled along the ray, it is not numerically identical to the final image error. The contribution tolerances determine how much spatial support of the coefficients is retained; the associated image differences and resource savings can then be quantified separately in benchmark cases.

Given a contribution tolerance $\epsilon_X$ for each coefficient and a set of representative times $\mathcal T_{\rm pilot}$ that samples changes in the radiative structure, let $\mathcal T_X\subseteq\mathcal T_{\rm pilot}$ be the set of times for which the denominator of Equation~\eqref{eq:region-support} is nonzero, and define the time-averaged contribution as
\begin{equation}
\overline C_X(\Mc R_k)
=
\frac{1}{\lvert\mathcal T_X\rvert}
\sum_{t\in\mathcal T_X}C_X(\Mc R_k;t).
\label{eq:mean-region-support}
\end{equation}
For each $X\in\mathcal X$, the regions are ranked by decreasing $\overline C_X(\Mc R_k)$ and added successively to $\Omega_X$ until
\begin{equation}
\sum_{\Mc R_k\subseteq\Omega_X}
\overline C_X(\Mc R_k)
\geq 1-\epsilon_X,
\qquad
\Omega_{\rm SL}=\bigcup_{X\in\mathcal X}\Omega_X.
\label{eq:region-selection}
\end{equation}
Within $\Omega_{\rm SL}$, time-dependent snapshots are read according to Equation~\eqref{eq:fluid-time}; in all other regions, the complete radiative-transfer path reads the fixed snapshot corresponding to $\base$.

Hybrid slow light substantially reduces the time-dependent spatial domain, but $\Omega_{\rm SL}$ may still contain a small number of samples with very long propagation delays. For each sample $i$, let $t_{a(i)}$ and $t_{b(i)}$ be the adjacent snapshot times that bracket $t_{{\rm fluid},i}$. We define the relative time range required by each slow-light region as
\begin{equation}
\begin{aligned}
W_k&=\bigcup_{i\in\Mc R_k\cap\Omega_{\rm SL}}
\left[t_{a(i)}-\base,t_{b(i)}-\base\right],\\
W_{\rm req}&=\bigcup_k W_k,
\end{aligned}
\label{eq:region-window}
\end{equation}
The set $W_{\rm req}$ may consist of multiple disconnected time intervals. To allow the production calculation to load snapshots sequentially over a continuous time range as the output time advances, CoportSL approximates the complete requirement $W_{\rm req}$ with a continuous effective window $W_q=[\Delta t_-(q),\Delta t_+(q)]$ for a target sample coverage $0<q\leq1$, where $\Delta t_-(q)$ and $\Delta t_+(q)$ are the earlier- and later-time endpoints of the window.

Let $\mathcal I_{\rm SL}$ be the index set of all discrete samples in $\Omega_{\rm SL}$ and $N_{\rm SL}$ their number. The sample coverage of a window $W$ is defined as
\begin{equation}
P_{\rm sam}(W)=
\frac{\#\{i\in\mathcal I_{\rm SL}:\Delta t_i\in W\}}
{N_{\rm SL}},
\label{eq:window-measure}
\end{equation}
where $\#\{\cdot\}$ denotes the cardinality of a set. The quantity $P_{\rm sam}$ assigns equal weight to each discrete sample in the slow-light region and therefore directly gives the fraction of samples whose delays fall within the window. The affine step and pixel weight enter the path-contribution integral in Equation~\eqref{eq:region-support}; the spatial selection and time window thus characterize, respectively, the spatial distribution of the transfer coefficients and the statistical distribution of the sampled delays. We define the $\mathrm p_q$ window $W_q$ as the shortest continuous interval satisfying $P_{\rm sam}(W_q)\geq q$, where $0<q\leq1$. To construct it, the $N_{\rm SL}$ relative delays are first sorted in ascending order. We then examine each candidate interval containing $\lceil qN_{\rm SL}\rceil$ adjacent sorted samples and choose the one with the smallest time span. For example, $\mathrm{p99}$ is the narrowest continuous interval containing at least $99\%$ of the discrete samples.

When the delay distribution has a concentrated main body and sparse long tails, $W_q$ shortens the span of simultaneously loaded snapshots by excluding a tail fraction smaller than $1-q$. For a sample with $\Delta t_i$ outside $W_q$, CoportSL clips its relative delay to the nearest window boundary and reconstructs the corresponding fluid state using Equation~\eqref{eq:interp}. This is the approximation introduced by the effective window relative to the complete delay range. Its error in the full-Stokes images is tested directly against the full-delay window. The continuous window also allows the production calculation to advance with the image time and load the required snapshots sequentially.

Let the available GRMHD snapshots span coordinate times $[t_{\rm in},t_{\rm out}]$. Once the effective window has been determined, the allowed base-time range of the production images satisfies
\begin{equation}
t_{\rm base}+W_q\subseteq[t_{\rm in},t_{\rm out}],
\label{eq:base-time-range}
\end{equation}
which determines the production output interval with complete access to the effective window.

\subsection{Algorithm Summary}\label{subsec:workflow}

The CoportSL procedure comprises three stages, as shown in Figure~\ref{fig:workflow}. Its inputs are a fixed observer, camera, and background spacetime; GRMHD snapshots ordered by coordinate time and sharing a fixed grid; a radiative model, the parameters needed for unit conversion, and a target observed flux for calibration; and the reference region $\Omega_{\rm ref}$, source-domain partition, representative-time set $\mathcal T_{\rm pilot}$, contribution tolerances $\epsilon_X$, and window coverage $q$.

The first stage prepares data shared by the slow-light calculation. A short, high-cadence sequence determines a snapshot interval that resolves changes in both the fluid and the observables. Unit conversion, flux calibration, and the electron distribution jointly provide the dimensional local quantities required to evaluate the radiative coefficients. The geodesics, parallel-transported polarization bases, spatial-interpolation information, and propagation delays depend only on the fixed geometry and grid and can therefore be reused throughout the image sequence. The reference region $\Omega_{\rm ref}$ then defines the zero point of the relative delay $\Delta t_i$ in Equation~\eqref{eq:time-offset}.

The second stage determines the spatial region and time window from representative times. The source partition defines the elementary regions $\Mc R_k$. The path-weighted contributions of the six transfer coefficients determine the slow-light region $\Omega_{\rm SL}$ through Equation~\eqref{eq:region-selection}, and the relative-delay distribution within that region determines $W_q$ through Equation~\eqref{eq:window-measure}. This sequence first identifies where fluid evolution must be retained and then determines which snapshots those regions must access simultaneously.

The third stage generates the production images and observables. As the base time advances, the calculation loads the GRMHD snapshots covered by $W_q$, reads the path-dependent fluid time inside $\Omega_{\rm SL}$ and the base time elsewhere, and integrates Equation~\eqref{eq:transfer} along the complete path. This yields full-Stokes images, fluxes, and polarization quantities. If a fast-light sequence is also generated, the effects of finite light-travel time on the observables can be compared in post-processing.

\section{CoportSL Benchmarks with M87*-like GRMHD Data}\label{sec:setup}

To test the accuracy and performance of the CoportSL slow-light calculation, we use an M87*-like GRMHD data set and examine a compact near-horizon emission region and an extended jet emission region, thereby testing the method as the source scale and propagation-delay range change.

\subsection{Numerical Setup}\label{subsec:grmhd}

\begin{figure*}[!t]
\centering
\includegraphics[width=0.90\textwidth]{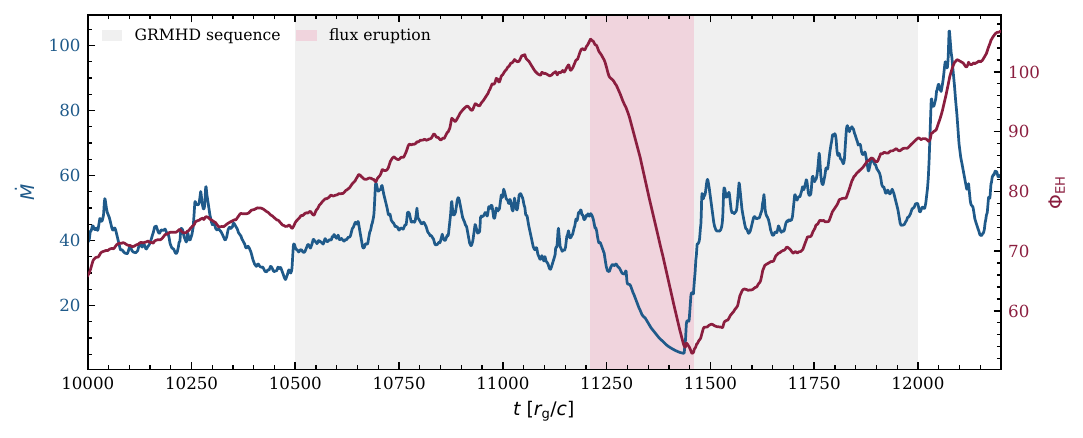}
\caption{Horizon accretion rate $\dot M$ and absolute event-horizon magnetic flux $\Phi_{\rm EH}$ in the GRMHD sequence, with $\dot M$ in code units. The light gray band marks the $10500$--$12000\,\tg$ interval used for radiative post-processing, and the pink band marks the third magnetic-flux eruption at $11210$--$11460\,\tg$ identified by \citet{Zhou_2026}. During the eruption, the magnetic flux decreases rapidly and the accretion rate reaches a trough, showing that the selected sequence contains appreciable nonstationary evolution.}
\label{fig:simulation-context}
\end{figure*}

\begin{table*}[!t]
\centering
\caption{Imaging setups for the near-horizon and jet images. The two image classes use the same GRMHD sequence, observer direction, and radiative-transfer outer boundary; only parameters that differ between the two setups are listed.}
\label{tab:imaging-settings}
\renewcommand{\arraystretch}{1.12}
\begin{tabular}{lll}
\toprule
\tblcell{0.24\textwidth}{Parameter}
& \tblcell{0.34\textwidth}{Near-horizon image}
& \tblcell{0.34\textwidth}{Jet image} \\
\midrule
\tblcell{0.24\textwidth}{Observing frequency}
& \tblcell{0.34\textwidth}{$230\,{\rm GHz}$}
& \tblcell{0.34\textwidth}{$86\,{\rm GHz}$} \\
\tblcell{0.24\textwidth}{Electron model}
& \tblcell{0.34\textwidth}{$\Mc T$, $\Mc P$}
& \tblcell{0.34\textwidth}{$\Mc P$} \\
\tblcell{0.24\textwidth}{Field of view}
& \tblcell{0.34\textwidth}{$\pi/64\,{\rm rad}$}
& \tblcell{0.34\textwidth}{$\pi/6\,{\rm rad}$} \\
\tblcell{0.24\textwidth}{Image resolution}
& \tblcell{0.34\textwidth}{$512^2$}
& \tblcell{0.34\textwidth}{$1024^2$} \\
\tblcell{0.24\textwidth}{Electron-number-density floor}
& \tblcell{0.34\textwidth}{$n_e\geq100\,{\rm cm}^{-3}$}
& \tblcell{0.34\textwidth}{$n_e\geq1\,{\rm cm}^{-3}$} \\
\bottomrule
\end{tabular}
\end{table*}

We use the M87*-like magnetically arrested disk (MAD) GRMHD sequence generated by \citet{Zhou_2026} \citep{Narayan2003}. The simulation was performed with the Black Hole Accretion Code (BHAC) in a fixed Kerr spacetime with dimensionless spin $a=0.9375$ \citep{Porth2017}, using modified Kerr--Schild (MKS) coordinates \citep{McKinney2004} and static mesh refinement (SMR), with an effective resolution of $384\times192\times256$. Each snapshot provides the primitive variables $\{\rho,u_{\rm g},\tilde u^i,\tilde B^i\}$, where $\rho$ and $u_{\rm g}$ are the rest-mass density and gas internal-energy density, and $\tilde u^i$ and $\tilde B^i$ are the spatial components of the four-velocity and magnetic field measured by an Eulerian observer, respectively. These quantities are first interpolated to the ray samples and then used to construct the fluid four-velocity $u^\mu$, comoving magnetic-field four-vector $b^\mu$, and local radiative-transfer coefficients. The density and internal energy in the strongly magnetized funnel are susceptible to the numerical floors used in GRMHD. We therefore evaluate the transfer coefficients only where $\sigma_{\rm M}=b^2/\rho\leq20$, following the imaging setup of \citet{Zhou_2026}. The GRMHD initial conditions, numerical evolution, and complete setup of the MAD magnetic-flux eruptions are described by \citet{Zhou_2026}.

To ensure that the benchmarks cover a stage of appreciable fluid evolution, we select a GRMHD interval containing the third magnetic-flux eruption identified by \citet{Zhou_2026}. The horizon accretion rate and absolute event-horizon magnetic flux used to characterize this event are defined as \citep{Tchekhovskoy:2011zx}
\begin{equation}
\begin{aligned}
\dot M&=-\int_{r=r_{\rm h}}\rho u^r\sqrt{-g}\,\ud\theta\ud\phi,\\
\Phi_{\rm EH}&=\frac{1}{2}\int_{r=r_{\rm h}}|\tilde B^r|\sqrt{\gamma}\,\ud\theta\ud\phi.
\end{aligned}
\end{equation}
where $r_{\rm h}$ is the event-horizon radius, $g$ and $\gamma$ are the determinants of the four-dimensional and spatial metrics, and $\tilde B^r$ is the radial magnetic-field component measured by an Eulerian observer in the $3+1$ decomposition. Figure~\ref{fig:simulation-context} shows their time evolution, with $\dot M$ in code units. The radiative post-processing inputs cover $t=10500$--$12000\,\tg$, with the third eruption at $11210$--$11460\,\tg$.

Single-fluid GRMHD data do not directly provide the electron temperature, so an electron-thermodynamic model must be introduced during radiative post-processing \citep{Mizuno:2021esc}. From the electron models of \citet{Zhou_2026}, we select the purely thermal model $\Mc T$ and the thermal plus isotropic nonthermal power-law model $\Mc P$, representing emission without and with a nonthermal high-energy tail, respectively. Nonthermal electrons can modify the spectrum, image structure, and extended jet emission of M87*-like sources \citep{Davelaar:2019jxr,Fromm:2021mqd,Zhang:2024ddt}. Both models determine the electron temperature using an $R$--$\beta$ prescription. In $\Mc P$, the nonthermal spectral index and energy fraction follow empirical relations from particle-in-cell (PIC) simulations of magnetic reconnection \citep{Monika2016,Ball_2018}. Their precise definitions are given in Appendix~\ref{app:electron-models}.

To convert the dimensionless length and time in the GRMHD simulation to the physical scales of M87* and determine the angular scale on the sky, we adopt a black hole mass $M=6.5\times10^9\,M_\odot$ and distance $D=16.9\,{\rm Mpc}$, consistent with existing observational constraints \citep{M87_4}. Here $M$ sets the physical scales of the gravitational radius $\rg$ and time $\tg$, while $D$ converts image-plane lengths to angular scales on the sky. Following Section~\ref{subsec:snapshot-cadence}, we calibrate the flux using the relatively quiescent interval $t=10500$--$11200\,\tg$ preceding the third eruption. Let $F_\nu$ be the total flux density obtained by integrating the image at observing frequency $\nu$. The calibration target is $\langle F_\nu\rangle=0.66\,{\rm Jy}$ at $230\,{\rm GHz}$ over this interval \citep{M87_4,Zhou_2026}. This yields $\dot M_{\Mc T}=3.7\times10^{-4}\,M_\odot\,{\rm yr}^{-1}$ and $\dot M_{\Mc P}=2.46\times10^{-4}\,M_\odot\,{\rm yr}^{-1}$. The nonthermal high-energy electrons in model $\Mc P$ increase the synchrotron emission per unit accretion rate, so the value of $\dot M_{\Mc P}$ required to reach the same target flux is lower than $\dot M_{\Mc T}$.

The observer is located at $(r,\theta,\phi)_{\rm obs}=(500\,\rg,163^\circ,10^{-4}\,{\rm rad})$, with the inclination following the imaging setup of \citet{Zhou_2026}. The ray-integration outer boundary $R_{\rm source}=200\,\rg$ encloses the millimeter emission region of the present GRMHD model and serves as the common complete source domain for the regional-contribution and selection tests. The fixed background, observer, and static grid allow all snapshots to share the geodesics and spatial-interpolation information. On top of these common settings, we construct near-horizon and jet images to compare the slow-light region and time-range requirements of compact and extended emission regions. The near-horizon images use a field of view of approximately $96\,\mu{\rm as}$ to resolve the black hole shadow and surrounding accretion-flow structure and use electron models $\Mc T$ and $\Mc P$ separately. The jet image expands the field of view to approximately $1.0\,{\rm mas}$, increases the image resolution, and lowers the electron-number-density floor used to evaluate the radiative coefficients so as to include the low-density extended jet. It uses model $\Mc P$ and retains the accretion-rate normalization obtained from its $230\,{\rm GHz}$ flux calibration, so the two image classes represent the same physical source. Table~\ref{tab:imaging-settings} summarizes the two imaging setups. Hereafter, we refer to them as the ``near-horizon image'' and ``jet image''; observing frequency is used only as an imaging parameter.

All radiative-transfer results reported here, including the data used in the accuracy tests and performance benchmarks, were generated with our CoportSL software. CoportSL also provides a Windows graphical user interface (GUI) that integrates flux calibration, slow-light pre-analysis, fast- and slow-light imaging, regional-error evaluation, post-processing, and performance benchmarking. The software implements the contribution-constrained hybrid slow-light method described in Section~\ref{sec:method} and is planned for further development into the general GRRT software Coport Studio. Its software architecture, extended capabilities, and independent validation will be presented separately by F. Zhou et al. (2026, in preparation).

\subsection{GRMHD Snapshot Cadence}\label{subsec:cadence-test}

\begin{figure*}[!t]
\centering
\includegraphics[width=0.88\textwidth]{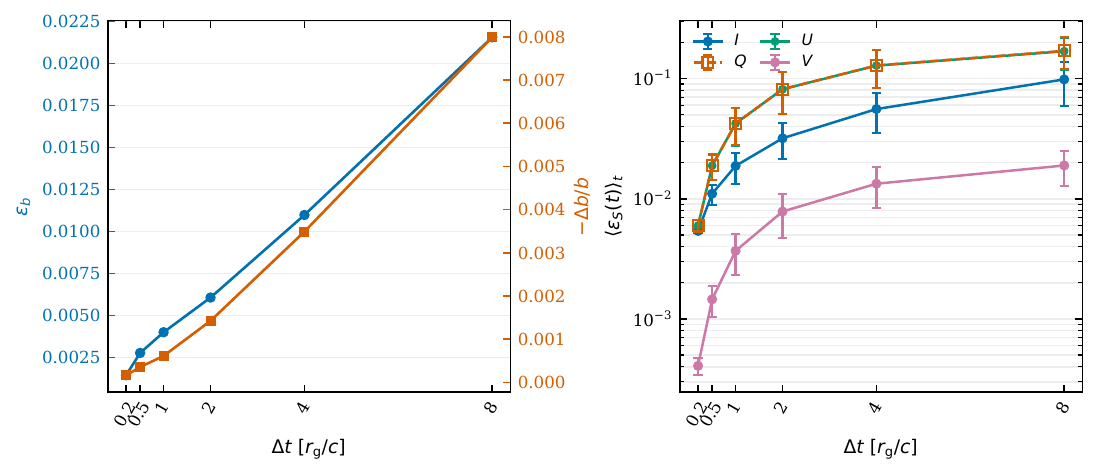}
\caption{GRMHD temporal-interpolation errors. The left panel shows the normalized absolute difference $\epsilon_b$ and signed bias $-\Delta b/b$ of the comoving magnetic-field strength. The right panel shows $\langle\epsilon_S(t)\rangle_t$ for Stokes $I/Q/U/V$ images relative to the reference sequence with $\Delta t_0=0.1\,\tg$. All image differences are normalized by the total Stokes $I$ of the reference image at the same time, and the error bars show the temporal standard deviation. The linear-polarization components $Q/U$ increase most rapidly with the sampling interval and therefore provide the main constraint on the adopted cadence.}
\label{fig:time-sampling}
\end{figure*}

\begin{table*}[!t]
\centering
\caption{Candidate slow-light regions used in the region scans. The notation $\Mc R_{i:j}$ denotes the union from $\Mc R_i$ through $\Mc R_j$. Near-horizon regions are named by their outer slow-light boundary, whereas the jet-image regions $\Omega_i$ successively expand the time-dependent domain in the north jet, south jet, and non-jet region. The largest region covering the complete source domain is denoted $\mathrm{Full}$ for both partitions.}
\label{tab:slow-regions}
\renewcommand{\arraystretch}{1.08}
\begin{tabular}{llll}
\toprule
\tblcell{0.105\textwidth}{Near-horizon name}
& \tblcell{0.285\textwidth}{Included elementary regions}
& \tblcell{0.105\textwidth}{Jet name}
& \tblcell{0.385\textwidth}{Included elementary regions} \\
\midrule
\tblcell{0.105\textwidth}{$\mathrm{r20}$} & \tblcell{0.285\textwidth}{$\Mc R_1$} & \tblcell{0.105\textwidth}{$\Omega_1$} & \tblcell{0.385\textwidth}{$\Mc R_1\cup\Mc R_7\cup\Mc R_{13:15}$} \\
\tblcell{0.105\textwidth}{$\mathrm{r30}$} & \tblcell{0.285\textwidth}{$\Mc R_{1:2}$} & \tblcell{0.105\textwidth}{$\Omega_2$} & \tblcell{0.385\textwidth}{$\Mc R_1\cup\Mc R_{7:8}\cup\Mc R_{13:15}$} \\
\tblcell{0.105\textwidth}{$\mathrm{r50}$} & \tblcell{0.285\textwidth}{$\Mc R_{1:3}$} & \tblcell{0.105\textwidth}{$\Omega_3$} & \tblcell{0.385\textwidth}{$\Mc R_1\cup\Mc R_{7:10}\cup\Mc R_{13:15}$} \\
\tblcell{0.105\textwidth}{$\mathrm{r80}$} & \tblcell{0.285\textwidth}{$\Mc R_{1:4}$} & \tblcell{0.105\textwidth}{$\Omega_4$} & \tblcell{0.385\textwidth}{$\Mc R_1\cup\Mc R_{7:15}$} \\
\tblcell{0.105\textwidth}{$\mathrm{r100}$} & \tblcell{0.285\textwidth}{$\Mc R_{1:5}$} & \tblcell{0.105\textwidth}{$\Omega_5$} & \tblcell{0.385\textwidth}{$\Mc R_{1:2}\cup\Mc R_{7:17}$} \\
\tblcell{0.105\textwidth}{$\mathrm{Full}$} & \tblcell{0.285\textwidth}{$\Mc R_{1:6}$} & \tblcell{0.105\textwidth}{$\Omega_6$} & \tblcell{0.385\textwidth}{$\Mc R_{1:4}\cup\Mc R_{7:17}$} \\
\tblcell{0.105\textwidth}{} & \tblcell{0.285\textwidth}{} & \tblcell{0.105\textwidth}{$\mathrm{Full}$} & \tblcell{0.385\textwidth}{$\Mc R_{1:18}$} \\
\bottomrule
\end{tabular}
\end{table*}

The production snapshot cadence is determined using the procedure in Section~\ref{subsec:snapshot-cadence}. We use a short, high-cadence reference sequence of 691 frames spanning $t=10800$--$10869\,\tg$ with interval $\Delta t_0=0.1\,\tg$, thin it to candidate intervals from $0.2$ to $8\,\tg$, and linearly reconstruct the primitive variables and Stokes images component by component at the omitted times for the fluid- and image-level tests, respectively.

The fluid-variable test covers $1.5<r<20\,\rg$ and divides this domain into the three radial shells $1.5$--$5\,\rg$, $5$--$10\,\rg$, and $10$--$20\,\rg$. Let $b_j$ and $\widetilde b_j$ be the comoving magnetic-field strengths at sample $j$ in the reference and linearly reconstructed sequences, respectively, where sample $j$ runs over the grid cells and reconstructed intermediate times in the three shells and $b_j=\sqrt{b^\mu b_\mu}$. We define
\begin{equation}
\epsilon_b=\frac{\sum_j|\widetilde b_j-b_j|}{\sum_j b_j},
\qquad
-\frac{\Delta b}{b}=-\frac{\sum_j(\widetilde b_j-b_j)}{\sum_j b_j}.
\label{eq:magnetic-interp-error}
\end{equation}
A positive value of the second quantity indicates that componentwise interpolation systematically underestimates the magnetic-field strength. At the image level, we use fast-light near-horizon images from electron model $\Mc T$ as the endpoints at adjacent retained times, linearly interpolate each Stokes component pixel by pixel, and compare the result with the high-cadence fast-light reference at the same time. A pixelwise relative error would be strongly amplified where the reference image is faint or a polarized component approaches zero because of the small local denominator, and would not represent the actual contribution of those pixels to the full-image difference. We therefore use the pixelwise absolute difference with a global normalization by the total Stokes $I$ of the reference image at the same time. For $S\in\{I,Q,U,V\}$, the normalized contribution of pixel $\boldsymbol{x}$ to the full-image difference is
\begin{equation}
\epsilon_S(\boldsymbol{x},t)=
\frac{\left|S(\boldsymbol{x},t)-S_{\rm ref}(\boldsymbol{x},t)\right|}
{\displaystyle\sum_{\boldsymbol{y}}
\left|I_{\rm ref}(\boldsymbol{y},t)\right|},
\label{eq:spatial-error}
\end{equation}
where $S$ and $S_{\rm ref}$ are the Stokes $S$ components of the image being compared and the reference image, and $I_{\rm ref}$ is the total intensity of the reference image. The denominator only makes the error dimensionless: $\epsilon_S(\boldsymbol{x},t)$ is not a relative error at that pixel, but its contribution to the full-image difference. Summing over all pixels gives the normalized full-image absolute difference,
\begin{equation}
\epsilon_S(t)=
\sum_{\boldsymbol{x}}\epsilon_S(\boldsymbol{x},t).
\label{eq:stokes-error}
\end{equation}
The quantity $\epsilon_S(t)$ measures the full-image absolute difference in component $S$ relative to the total Stokes $I$ of the reference image, and $\langle\epsilon_S(t)\rangle_t$ denotes its time average over the comparison interval. All subsequent image-accuracy tests use these definitions.

Figure~\ref{fig:time-sampling} shows that the interpolation errors in both the fluid variables and the Stokes images increase with the candidate interval. Stokes $Q/U$ encode both the amplitude and direction of linear polarization and are more sensitive to the reconstructed magnetic-field direction and polarized-transfer coefficients, so their differences exceed those in $I/V$. We adopt $\Delta t=0.5\,\tg$ for the production calculations. At this interval, $\epsilon_b=2.77\times10^{-3}$ and the signed bias in the magnetic-field strength is $3.55\times10^{-4}$, indicating an overall underestimate by the reconstruction. The time-averaged differences in Stokes $I$, $Q$, $U$, and $V$ are $1.10\times10^{-2}$, $1.88\times10^{-2}$, $1.88\times10^{-2}$, and $1.45\times10^{-3}$, respectively. A shorter interval would reduce the interpolation error but would also increase the storage required by the long sequence and increase, approximately inversely with the interval, the number of snapshot layers loaded simultaneously within a fixed delay window. The choice $0.5\,\tg$ is therefore the compromise adopted here between temporal-discretization accuracy and storage and memory costs. The subsequent spatial-selection and time-window tests use the same interval, allowing the image differences introduced by the spatial region and time window to be quantified separately.

\subsection{Slow-light Region and Effective Time Window}\label{subsec:delay-window}

\begin{figure*}[!t]
\centering
\includegraphics[width=0.92\textwidth]{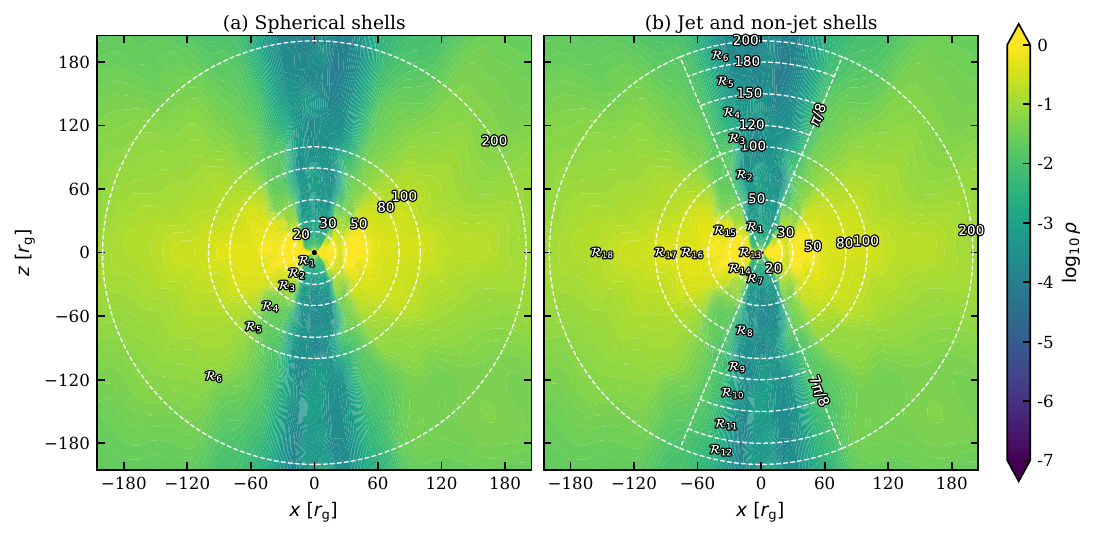}
\caption{Elementary regions for the near-horizon and jet images. The background shows the simulated density $\log_{10}\rho$ in the $xz$ plane at $t=10800\,\tg$; white dashed lines mark region boundaries and numbers give $r/\rg$. The left panel partitions the complete source domain into six radial shells. In the right panel, $\theta=\pi/8$ and $7\pi/8$ separate the north jet, south jet, and non-jet region, each of which is divided radially. The two partitions describe the coefficient distributions in the near-horizon accretion flow and extended jet, respectively.}
\label{fig:region-partition}
\end{figure*}

\begin{figure*}[!t]
\centering
\includegraphics[width=0.94\textwidth]{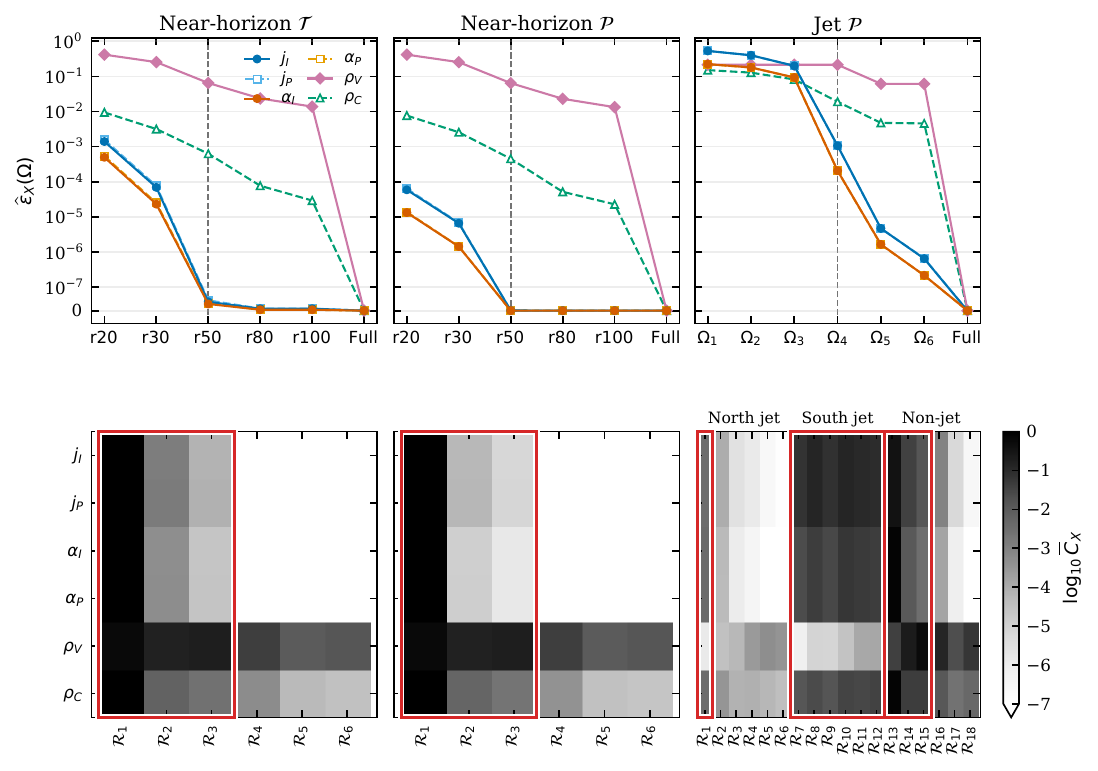}
\caption{Actual omitted fractions of the candidate slow-light regions and time-averaged coefficient contributions of the elementary regions. The three columns show model $\Mc T$ for the near-horizon image, model $\Mc P$ for the near-horizon image, and model $\Mc P$ for the jet image, at observing frequencies of $230\,{\rm GHz}$, $230\,{\rm GHz}$, and $86\,{\rm GHz}$, respectively. The top row shows the fractions of coefficient contribution not covered by each candidate slow-light region, $\widehat\epsilon_X(\Omega)$; vertical dashed lines mark the regions adopted for the demonstration. In the bottom row, grayscale represents $\log_{10}\overline C_X(\Mc R_k)$ for the six transfer coefficients in the elementary regions, and red borders mark the elementary regions obtained from Equation~\eqref{eq:region-selection}. In the near-horizon images, emission, absorption, and Faraday conversion are concentrated within $r<50\,\rg$, while Faraday rotation extends to larger radii. In the jet image, both emission and propagation coefficients extend along the south jet, and the non-jet region also provides appreciable Faraday contributions.}
\label{fig:preanalysis-coeff-distribution}
\end{figure*}

\begin{figure*}[!t]
\centering
\includegraphics[width=0.92\textwidth]{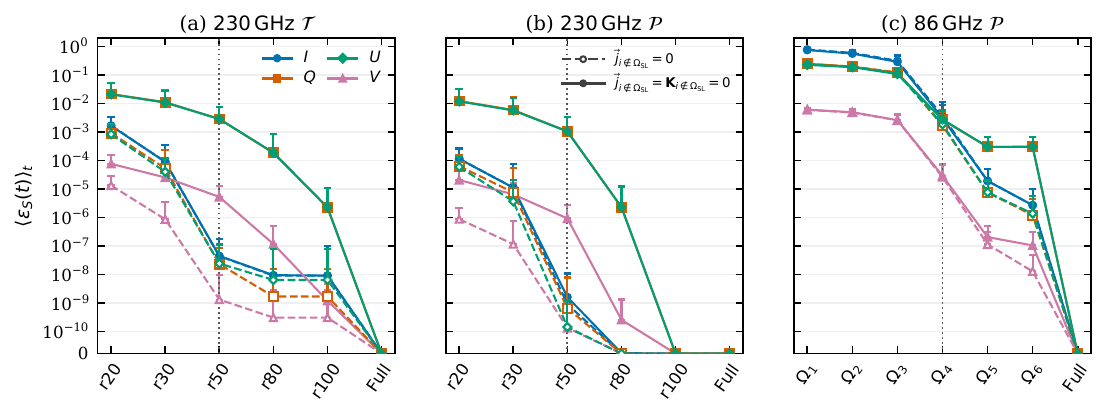}
\caption{Tests that suppress radiative coefficients outside each candidate region. Dashed curves suppress only the outer-region emission vector $\vec j$, whereas solid curves suppress both the emission vector and propagation matrix $\mathbf K$. The vertical axis is $\langle\epsilon_S(t)\rangle_t$ relative to the image retaining all coefficients, as defined in Equation~\eqref{eq:stokes-error}. The three columns show model $\Mc T$ for the near-horizon image, model $\Mc P$ for the near-horizon image, and model $\Mc P$ for the jet image; vertical dash-dotted lines mark the production selections. Direct outer-region emission is negligible in the near-horizon images, although propagation terms still affect $Q/U$. For the jet image, the difference decreases rapidly near $\Omega_4$, supporting the jet--non-jet combination selected by the contribution analysis.}
\label{fig:region-error}
\end{figure*}

\begin{figure*}[!t]
\centering
\includegraphics[width=0.868\textwidth]{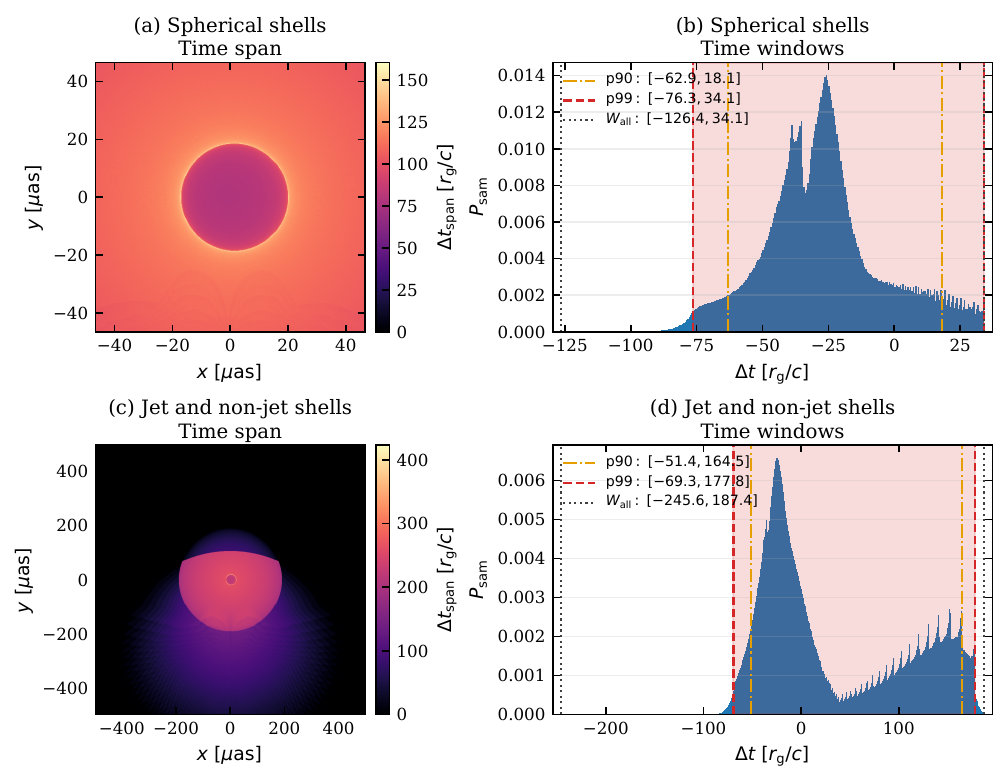}
\caption{Geometric delays in the two production slow-light selections. The top row corresponds to the near-horizon shell selection $\Mc R_{1:3}$, and the bottom row to the jet--non-jet selection $\Omega_4$. The left column shows, for each pixel, the difference $\Delta t_{\rm span}$ between the maximum and minimum delays of samples on its ray within the selected region. The right column shows the relative-delay distribution of the samples within the selected region and the coverage windows defined by Equation~\eqref{eq:window-measure}. Dash-dotted, dashed, and dotted lines mark $\mathrm{p90}$, $\mathrm{p99}$, and $W_{\rm all}$, which covers all samples, respectively; pink shading marks the adopted $\mathrm{p99}$ window. The jet selection has both a larger within-ray delay span and a broader $\mathrm{p99}$ window and therefore requires more GRMHD snapshot layers than the near-horizon setup.}
\label{fig:delay-analysis}
\end{figure*}

Constructing the slow-light region first requires partitioning the complete source domain and combining elementary regions into candidate slow-light regions for comparing the selection extent. Following Section~\ref{subsec:region-general}, $\Mc R_k$ denotes an elementary region used to evaluate coefficient contributions, and we define $\Mc R_{i:j}\equiv\bigcup_{k=i}^{j}\Mc R_k$. For the near-horizon images, the complete source domain is partitioned into radial shells $\Mc R_k=\{r_k<r<r_{k+1}\}$. When the slow-light region comprises consecutive shells $\Mc R_{1:m}$ from the center outward, its outer boundary defines the hybrid slow-light radius $\slow=r_{m+1}$, such that $\Omega_{\rm SL}=\Mc R_{1:m}=\{r<\slow\}$. For the jet image, conical surfaces at $\theta=\pi/8$ and $7\pi/8$, placed along the edges of the low-density funnels, separate the north jet, south jet, and non-jet region. Each of these three parts is divided radially to resolve the coefficient distributions in the extended jets and compact accretion flow. Figure~\ref{fig:region-partition} shows the geometry and numbering of the two elementary-region partitions, and Table~\ref{tab:slow-regions} lists the combinations used as candidate slow-light regions.

The regional-contribution calculation uses the common GRMHD and radiative-transfer settings of Section~\ref{subsec:grmhd}. It samples every $10\,\tg$ over $t=10500$--$12000\,\tg$, for a total of 151 representative snapshots. For any candidate slow-light region $\Omega$, we define the fraction of coefficient contribution not covered by that region as
\begin{equation}
\widehat\epsilon_X(\Omega)
=1-\sum_{\Mc R_k\subseteq\Omega}\overline C_X(\Mc R_k).
\label{eq:effective-region-tolerance}
\end{equation}
Hereafter, we call $\widehat\epsilon_X(\Omega)$ the actual omitted fraction of the candidate region. It is evaluated after the candidate region has been fixed and differs from the contribution tolerance $\epsilon_X$ prescribed in Equation~\eqref{eq:region-selection} to control the selection extent. The top row of Figure~\ref{fig:preanalysis-coeff-distribution} shows the actual omitted fractions as the candidate slow-light region expands, while the bottom row shows the time-averaged contributions of the elementary regions.

The two electron models used for the near-horizon images have similar radial contribution distributions. The coefficients $j_I$, $j_P$, $\alpha_I$, and $\alpha_P$ are strongly concentrated in $\Mc R_{1:3}$, whose cumulative contribution exceeds $1-10^{-7}$ for all four coefficients. For models $\Mc T$ and $\Mc P$, this region contains approximately $99.94\%$ and $99.96\%$ of the $\rho_C$ contribution, respectively, and approximately $93.5\%$ of the $\rho_V$ contribution in both cases. The near-horizon radiation is therefore emitted and absorbed predominantly in the inner region, while Faraday rotation outside it can continue to alter the linear polarization after the rays leave that region.

For the jet image, the north jet, south jet, and non-jet region correspond to $\Mc R_{1:6}$, $\Mc R_{7:12}$, and $\Mc R_{13:18}$, respectively. The observer lies in the southern hemisphere at $\theta=163^\circ$, approximately $17^\circ$ from the south-jet axis, so the south jet is approaching. Relativistic beaming enhances its emission and makes the receding north-jet emission comparatively weaker \citep{Tsunetoe:2025slowlight}, accounting for the north--south asymmetry in the emission and absorption contributions. The Faraday terms also have appreciable contributions in the inner non-jet region, so the selection must include elementary regions in both the jets and the non-jet region.

We use these two image classes to demonstrate how a slow-light region is constructed from coefficient contributions. The top row of Figure~\ref{fig:preanalysis-coeff-distribution} shows that, for the near-horizon images, the actual omitted fractions have already decreased markedly when the candidate region reaches $\mathrm{r50}$: the omitted fractions of emission and absorption are below $10^{-7}$, that of Faraday conversion is below $10^{-3}$, and that of Faraday rotation is approximately $6.5\%$. For the jet image, as the candidate region expands from $\Omega_3$ to $\Omega_4$, the omitted emission fraction drops from approximately $20\%$ to the $10^{-3}$ level, while the omitted fractions of the other coefficients also fall below their values for the preceding candidate region. Based on these changes, we adopt $\mathrm{r50}$ and $\Omega_4$ as the demonstration regions.

These demonstration regions must be converted into reproducible selection conditions through the contribution tolerances. The candidate regions in Table~\ref{tab:slow-regions} display how cumulative contributions change as the spatial extent expands; the actual selection still ranks the elementary regions for each coefficient and takes their union according to Equation~\eqref{eq:region-selection}. To reproduce the demonstration regions in the present cases, we set each contribution tolerance by reference to the actual omitted fractions of adjacent candidate regions. For a coefficient that determines whether the final elementary region is included, $\epsilon_X$ is placed between the actual omitted fractions of the demonstration region and the preceding candidate region, causing the selection to include the regions before the conspicuous drop in contribution. Ordered as $(j_I,j_P,\alpha_I,\alpha_P,\rho_V,\rho_C)$, the near-horizon tolerances are $(10^{-3},10^{-3},10^{-3},10^{-3},0.1,2\times10^{-3})$, and the jet-image tolerances are $(2\times10^{-3},2\times10^{-3},10^{-3},10^{-3},0.26,0.025)$. Equation~\eqref{eq:region-selection} then gives $\Mc R_{1:3}$, or $\mathrm{r50}$, for both near-horizon electron models and $\Mc R_1\cup\Mc R_{7:15}$, or $\Omega_4$, for the jet image. The latter contains 10 elementary regions, covering the south jet, the inner north jet, and the inner non-jet region.

The regional contributions determine only where fluid evolution is retained, not whether radiative transfer at that location is retained. The production calculation fixes the fluid time outside the selected region but still evaluates all transfer coefficients along the complete ray. To show why the radiative processes outside the selection cannot also be discarded, Figure~\ref{fig:region-error} deliberately applies stronger suppression approximations: dashed curves set only the emission vector $\vec j$ to zero outside each candidate region, whereas solid curves set both $\vec j$ and the propagation matrix $\mathbf K$ to zero.

For the production $\mathrm{r50}$ setting of the near-horizon images, suppressing outer-region emission yields time-averaged differences no larger than $4.37\times10^{-8}$ in any Stokes component. When propagation is also suppressed, the $Q/U$ differences are approximately $2.85\times10^{-3}$ and $1.05\times10^{-3}$ for models $\Mc T$ and $\Mc P$, respectively. Direct emission from the outer region is negligible, but absorption and Faraday propagation still modify the polarization. The production hybrid scheme therefore freezes the outer-region fluid time while retaining all transfer coefficients along the complete path.

For the jet image, the $\Omega_4$ setting yields $I/Q/U$ differences of several $10^{-3}$ under both suppression approximations. Suppressing only outer-region emission gives a Stokes $I$ difference of $3.70\times10^{-3}$, while additionally suppressing propagation gives $Q/U$ differences of $2.66\times10^{-3}$ and $2.75\times10^{-3}$. The differences produced by the two approximations do not add linearly, showing that signed polarized emission and propagation effects can partially cancel in the image. Both outer-region emission and propagation affect the image; the production calculation therefore fixes the fluid time outside $\Omega_4$ but does not truncate radiative transfer.

After the spatial region has been selected, the time window is constructed from the geometric-delay distribution of the discrete samples within that region. We denote the continuous window that covers all delays in the selected region, corresponding to $q=1$, by $W_{\rm all}$. Both partitions contain the compact source region $r<20\,\rg$, which we use as their common reference region $\Omega_{\rm ref}$ so that the delay zero point in Equation~\eqref{eq:time-offset} is identical in the two analyses. The two near-horizon electron models select the same $\Mc R_{1:3}$ and therefore share the shell-delay results in the top row of Figure~\ref{fig:delay-analysis}. All samples span $[-126.4,34.1]\,\rg/c$, while the $\mathrm{p99}$ window is $[-76.3,34.1]\,\rg/c$, corresponding to 223 consecutive snapshot layers at $\Delta t=0.5\,\tg$.

The bottom row of Figure~\ref{fig:delay-analysis} shows the delay results for the selected jet--non-jet region. All samples span $[-245.6,187.4]\,\rg/c$, with a maximum span of $423.1\,\rg/c$. The $\mathrm{p99}$ window is $[-69.3,177.8]\,\rg/c$, has a width of $247.1\,\rg/c$, and corresponds to 496 snapshot layers. By comparison, the near-horizon $\mathrm{p99}$ window has a width of $110.4\,\rg/c$ and requires only 223 layers. The signs of the window endpoints depend on the reference-zero convention in Equation~\eqref{eq:time-offset}; the snapshot requirement is controlled by the window width, not by the signs of its endpoints. The production near-horizon images therefore use the $\mathrm{r50}$ region and a $\mathrm{p99}$ window containing 223 snapshot layers, while the jet image uses $\Omega_4$ and a $\mathrm{p99}$ window containing 496 layers. Because the jet emission region is more extended and covers a broader geometric-delay range along the rays, it requires more snapshot layers to be retained simultaneously.

\subsection{Error Convergence Tests}\label{subsec:validation-scan}

\begin{figure*}[!t]
\centering
\includegraphics[width=0.94\textwidth]{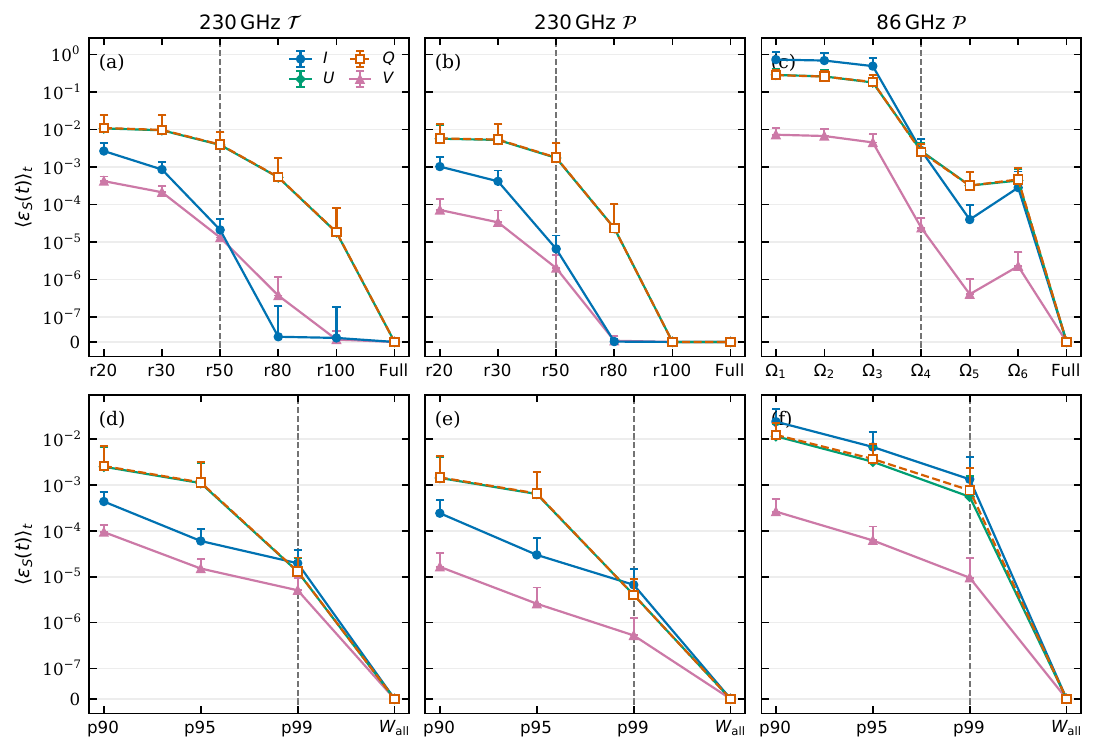}
\caption{Image-error convergence of the spatial selection and time-window approximations. The three columns show model $\Mc T$ for the near-horizon image, model $\Mc P$ for the near-horizon image, and model $\Mc P$ for the jet image. The top row fixes the sample coverage at $q=0.99$, constructs the corresponding $\mathrm{p99}$ window for each candidate region using Equation~\eqref{eq:window-measure}, and uses the $\mathrm{Full}$ region as the reference. The bottom row fixes the demonstration slow-light region, increases the window coverage successively, and uses $W_{\rm all}$ with $q=1$ as the reference. The image error is the time average of Equation~\eqref{eq:stokes-error}, error bars show the temporal standard deviation, and vertical dashed lines mark the demonstration settings. Under the same window-construction rule, the region errors decrease overall together with the actual omitted fractions in the top row of Figure~\ref{fig:preanalysis-coeff-distribution}; the jet image shows a local increase as the region expands further.}
\label{fig:validation-scan}
\end{figure*}

Section~\ref{subsec:delay-window} obtained the demonstration spatial regions and time windows from the coefficient contributions and delay coverage. We now expand the candidate slow-light region and increase the window coverage separately to test whether the Stokes images converge toward complete references as the two approximations are relaxed, quantify the image differences of the production settings, and compare the error trends with those inferred from the preceding contribution analysis and delay coverage. Figure~\ref{fig:validation-scan} uses the error defined in Equation~\eqref{eq:stokes-error} to scan the candidate region and window coverage separately. The region scan fixes the coverage at $q=0.99$, constructs a separate $\mathrm{p99}$ window for every candidate region using Equation~\eqref{eq:window-measure}, and uses the $\mathrm{Full}$ setting containing all elementary regions as the reference. Thus, the window-construction rule is fixed, rather than the same numerical time interval. The window scan fixes the demonstration slow-light region and compares $\mathrm{p90}$, $\mathrm{p95}$, $\mathrm{p99}$, and $W_{\rm all}$. All settings along a given scan are compared directly at common GRMHD times: only the candidate region changes in the top row, and only the window coverage changes in the bottom row. For each near-horizon model, the region scan contains 2253 frames and the window scan 2679 frames. The corresponding jet-image scans contain 2260 and 2134 frames, respectively.

As the near-horizon region expands, the actual omitted fraction of Faraday rotation in the top row of Figure~\ref{fig:preanalysis-coeff-distribution} and the linear-polarization errors in the top row of Figure~\ref{fig:validation-scan} follow the same downward trend. For model $\Mc T$, as the region expands from $\mathrm{r20}$ through $\mathrm{r50}$ to $\mathrm{r100}$, $\widehat\epsilon_{\rho_V}$ decreases from $0.415$ to $0.0647$ and $0.0138$, while the maximum Stokes difference decreases from $1.10\times10^{-2}$ to $3.98\times10^{-3}$ and $1.84\times10^{-5}$. For model $\Mc P$, the corresponding omitted fractions are $0.414$, $0.0641$, and $0.0133$, and the maximum differences are $5.83\times10^{-3}$, $1.80\times10^{-3}$, and $2.94\times10^{-10}$. The time-averaged Stokes $I$ differences between the production $\mathrm{r50}$ region and $\mathrm{Full}$ are $2.11\times10^{-5}$ and $6.57\times10^{-6}$ for models $\Mc T$ and $\Mc P$, respectively. The largest difference for each region occurs in $Q/U$, whereas the omitted emission and absorption fractions have fallen to $10^{-8}$ or below by $\mathrm{r50}$. This correspondence shows that the coefficient contributions identify outer-region Faraday rotation as the principal accuracy constraint for the near-horizon images.

For model $\Mc P$ in the jet image, the omitted emission fraction and the Stokes $I$ error drop sharply at the same candidate region. As the selection expands from $\Omega_1$ through $\Omega_3$ to the demonstration region $\Omega_4$, $\widehat\epsilon_{j_I}$ decreases from $0.529$ and $0.199$ to $1.06\times10^{-3}$, while the Stokes $I$ difference decreases from $0.736$ and $0.494$ to $2.76\times10^{-3}$. After further expansion to $\Omega_5$, the two quantities fall to $4.71\times10^{-6}$ and $3.97\times10^{-5}$, respectively. This result is of the same order as the $3.70\times10^{-3}$ difference obtained by suppressing emission outside $\Omega_4$ in Figure~\ref{fig:region-error}; together, the two tests identify extended emission as the principal accuracy constraint for the jet selection. From $\Omega_5$ to $\Omega_6$, however, the actual omitted fractions of all coefficients continue to decrease while the Stokes $I/Q/U$ differences increase locally. This nonmonotonicity may arise from cancellation between signed emission and propagation effects along the ray, whereas $\widehat\epsilon_X(\Omega)$ measures only the spatial fraction of the absolute local coefficient amplitude and retains neither sign nor propagation coupling. The actual omitted fraction is therefore not a pointwise numerical error bound for the image; it agrees with the image error only in the overall downward trend before the demonstration region and in the dominant component.

The window-coverage scan further establishes the image-error basis for adopting $\mathrm{p99}$. At $\mathrm{p95}$, the maximum component difference is already below $1.15\times10^{-3}$ for the near-horizon images but remains $6.77\times10^{-3}$ for the jet image. After the coverage is increased to $\mathrm{p99}$, the maximum window difference is no larger than $1.34\times10^{-3}$ in any of the three settings. Thus, among $\mathrm{p90}$, $\mathrm{p95}$, and $\mathrm{p99}$ in the present scan, $\mathrm{p99}$ is the first coverage for which the maximum window difference is below $4\times10^{-3}$ in all three settings. Relative to $W_{\rm all}$, the maximum component differences of the production $\mathrm{p99}$ window are $1.96\times10^{-5}$ and $6.65\times10^{-6}$ for near-horizon models $\Mc T$ and $\Mc P$, respectively, both in Stokes $I$. The time window modifies only the small number of delay samples outside $\mathrm{p99}$, so the ordering of the component sensitivities here need not match that in Section~\ref{subsec:cadence-test}, where all samples undergo temporal interpolation. For both near-horizon electron models, the images converge stably toward their complete references as the spatial region expands and the time window widens. Across the three production settings, the maximum image difference introduced independently by either the slow-light region or the time window remains below $4\times10^{-3}$. This statement quantifies the two approximations separately and does not represent a total error obtained by combining them with the temporal-interpolation error.

\subsection{Computational Performance and Memory Requirements}\label{sec:performance}

\begin{figure*}[!t]
\centering
\includegraphics[width=0.92\textwidth]{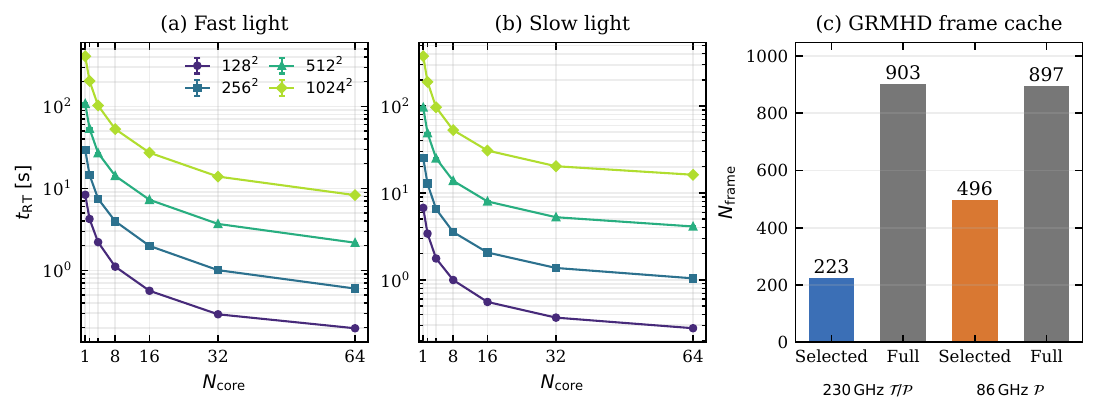}
\caption{Computational performance and GRMHD snapshot requirements of CoportSL. The first two panels use near-horizon model $\Mc T$ and show the per-frame radiative-transfer time $t_{\rm RT}$ for fast light and hybrid slow light with the production $\mathrm{r50}$ region and $\mathrm{p99}$ window, respectively. The vertical axes are logarithmic, and the legend entries from $128^2$ to $1024^2$ denote image resolutions from 128 to 1024 pixels per side. The third panel compares the snapshot-layer counts required by the production region plus $\mathrm{p99}$ window (``Selected'') and by full slow light. Models $\Mc T$ and $\Mc P$ share the near-horizon snapshot requirement, while the jet image uses model $\Mc P$. Both calculations become faster as the number of physical cores increases, although the time reduction for slow light becomes more gradual at high core counts. The spatial selection and effective time window reduce the snapshot-layer counts by $75.3\%$ and $44.7\%$ for the near-horizon and jet images, respectively.}
\label{fig:benchmark}
\end{figure*}

After establishing image convergence, we evaluate separately the per-frame computational performance and the data requirements of the production settings for long sequences. The performance benchmarks are run on an AMD EPYC 7702. The first two panels of Figure~\ref{fig:benchmark} compare the per-frame times of fast light and the production hybrid slow light, while the third panel and the following memory estimates compare the data requirements of the production settings and full slow light. The timing tests use near-horizon model $\Mc T$ throughout; the hybrid slow-light calculation uses the production $\mathrm{r50}$ region and $\mathrm{p99}$ window, while only the number of physical cores and image resolution vary. The tests cover 1, 2, 4, 8, 16, 32, and 64 physical cores and four resolutions from $128^2$ to $1024^2$. For each combination, one frame is used for warm-up, followed by a scan over consecutive output times. The means and standard deviations in the figure are calculated from the individual frame times over this interval, not from multiple independent batches. The per-frame radiative-transfer time $t_{\rm RT}$ measures the transfer stage after the geometric cache has been prepared and excludes the separately recorded primitive-variable frame update. For an image sequence that reuses the same geometric cache, this is the repeated computational cost at each output time.

At low core counts, fast and hybrid slow light have essentially comparable per-frame times, and both decrease substantially as the core count increases. For example, from 1 to 64 cores, the fast-light time for a $512^2$ image decreases from $107.7\,{\rm s}$ to $2.17\,{\rm s}$, whereas the slow-light time decreases from $96.6\,{\rm s}$ to $4.11\,{\rm s}$. At $1024^2$ resolution on 64 cores, the corresponding times are $8.27\,{\rm s}$ and $16.2\,{\rm s}$. In the present timing, the hybrid slow-light per-frame radiative-transfer time does not exceed that of fast light on one core and is approximately $1.9$ times the fast-light value on 64 cores. Overall, hybrid slow light remains comparable to fast light at low core counts and continues to accelerate as more cores are used, with more gradual gains at high core counts.

The more substantial resource saving comes from the number of GRMHD snapshot layers required simultaneously by one image and their memory consumption. Full slow light over the complete spatial domain and delay range requires 903 layers for the near-horizon images, whereas the production region plus $\mathrm{p99}$ window requires only 223, a reduction of $75.3\%$. For the jet image, the corresponding count decreases from 897 to 496, a reduction of $44.7\%$. The difference between these reductions is consistent with the delay distributions in Figure~\ref{fig:delay-analysis}: the extended jet has a broader $\mathrm{p99}$ window, so the production calculation must retain more snapshot layers. Snapshot-cache capacity is set by the fluid-grid size, number of primitive variables, and snapshot-layer count and is independent of image resolution; image resolution affects only per-pixel structures such as the ray cache. Based on the principal data structures in the source code---the primitive-variable snapshots, static grid, ray geometry, spatial-interpolation positions, region masks, and output images---full slow light for a $512^2$ near-horizon image requires approximately $65.7\,{\rm GiB}$, whereas the production region and $\mathrm{p99}$ window reduce this capacity to approximately $20.2\,{\rm GiB}$, a reduction of $69.2\%$. Models $\Mc T$ and $\Mc P$ share the same geometry, region, and time window and therefore have the same memory requirement. For the $1024^2$ jet image, the corresponding memory decreases from approximately $64.1\,{\rm GiB}$ to $37.3\,{\rm GiB}$, a reduction of $41.9\%$. Thus, the spatial selection and effective time window substantially reduce the capacity of the principal data structures in both image classes, with the larger reduction obtained for the compact near-horizon emission region.

As a cache-structure comparison, we use a fixed public version of the ipole slow-light code \citep{Moscibrodzka:2017lcu,Moscibrodzka:2021} (\href{https://github.com/AFD-Illinois/ipole/commit/7f7a482cf91125aeeeb9c431485bba680e8941d7}{commit \texttt{7f7a482}}) and connect it to the same BHAC data. The ipole code stores a complete trajectory for every pixel together with multiple unfinished image states and therefore has a large memory requirement. For near-horizon imaging, the principal caches of ipole and CoportSL require approximately $255\,{\rm GiB}$ and $20.2\,{\rm GiB}$, respectively, a reduction of $92.1\%$ for CoportSL. For jet imaging, the corresponding values are approximately $657\,{\rm GiB}$ and $37.3\,{\rm GiB}$, a reduction of $94.3\%$.

Because full slow light with ipole has a large memory requirement, we first measure its fast-light per-frame time on the same hardware and then combine this benchmark with low-resolution slow-light tests to estimate its mean per-frame time; all timings exclude snapshot input and result output. The mean ipole fast-light time is $96.4\,{\rm s}$, of which approximately $30\%$ is spent recomputing geodesics for each image; the estimated slow-light radiative-transfer time is approximately $38\,{\rm s}$ per frame. The CoportSL fast-light time on the same hardware is $1.65\,{\rm s}$, and Figure~\ref{fig:benchmark} shows comparable fast- and slow-light times at low core counts. The estimated mean per-frame reduction relative to ipole is therefore approximately $96\%$.

Overall, the spatial selection and time window reduce both the snapshot-layer count and the capacity of the principal data structures for the two image classes, while hybrid slow light retains a per-frame radiative-transfer time comparable to fast light at low core counts. The compact near-horizon emission region has smaller spatial and delay extents and therefore yields larger resource savings than the extended jet.

\section{Comparison of Fast- and Slow-light Observables}\label{sec:results}

Having established the image differences and resource requirements of CoportSL, we now compare fast- and slow-light results from the same GRMHD sequence to examine the effects of finite light-travel time on the total flux and full-Stokes spatial structure.

\subsection{Time Alignment and Differences of the Total-flux Curves}\label{subsec:time-alignment}

The numerical label $\base$ of a slow-light image is the reference fluid time defined in Section~\ref{subsec:slow-scheme}; it does not imply that all radiation in the image originates at the same fluid time. If the dominant emission locations have approximately the same propagation delay, slow light appears in the total flux primarily as a global time offset, and the shifted fast- and slow-light curves can retain the same evolutionary form \citep{Bronzwaer:2018}. Conversely, if the propagation delay varies appreciably with emission location, a slow-light image combines radiation from different fluid times, and a single global time shift generally recovers only part of its spatially integrated variability. We use the peak of the total-flux cross-correlation to test whether a global shift can establish a common time coordinate for the two sequences, and define the alignment time of the slow-light sequence as
\begin{equation}
t_{\rm align}=t_{\rm base}+\Delta t_{\rm align},
\label{eq:aligned-time}
\end{equation}
where the fast-light sequence uses $t_{\rm align}=t$, and $\Delta t_{\rm align}$ is the global shift added to the numerical slow-light label. It shifts the slow-light curve as a whole and differs from the sample-level relative delays used in Section~\ref{subsec:delay-window} to construct the snapshot window of a single image. We search time shifts over $[-300,300]\,\rg/c$ in steps of $0.5\,\rg/c$. For each candidate shift, only data points in the overlap of the two sequences are compared, and the number of overlapping points must be at least $75\%$ of the number shared by the unshifted sequences. The shift that maximizes the Pearson correlation coefficient is selected.

After time alignment, we compare the two curves using their correlation coefficient, mean-flux difference, and normalized root-mean-square residual. The correlation coefficient measures whether the shifted curves vary synchronously, the mean-flux difference measures their overall emission levels, and the rms residual measures the amplitude of their point-by-point differences. We normalize the rms residual by the mean slow-light flux and define
\begin{equation}
\delta_{F,{\rm rms}}=
\frac{\left\langle\left[F_{\nu,{\rm fast}}(t_{\rm align})-F_{\nu,{\rm slow}}(t_{\rm align})\right]^2\right\rangle_{t_{\rm align}}^{1/2}}
{\left\langle F_{\nu,{\rm slow}}(t_{\rm align})\right\rangle_{t_{\rm align}}}.
\label{eq:flux-rms}
\end{equation}
All fast- and slow-light curves below are compared over $t_{\rm align}=10800$--$11600\,\tg$.

\begin{figure*}[t]
\centering
\includegraphics[width=0.91\textwidth]{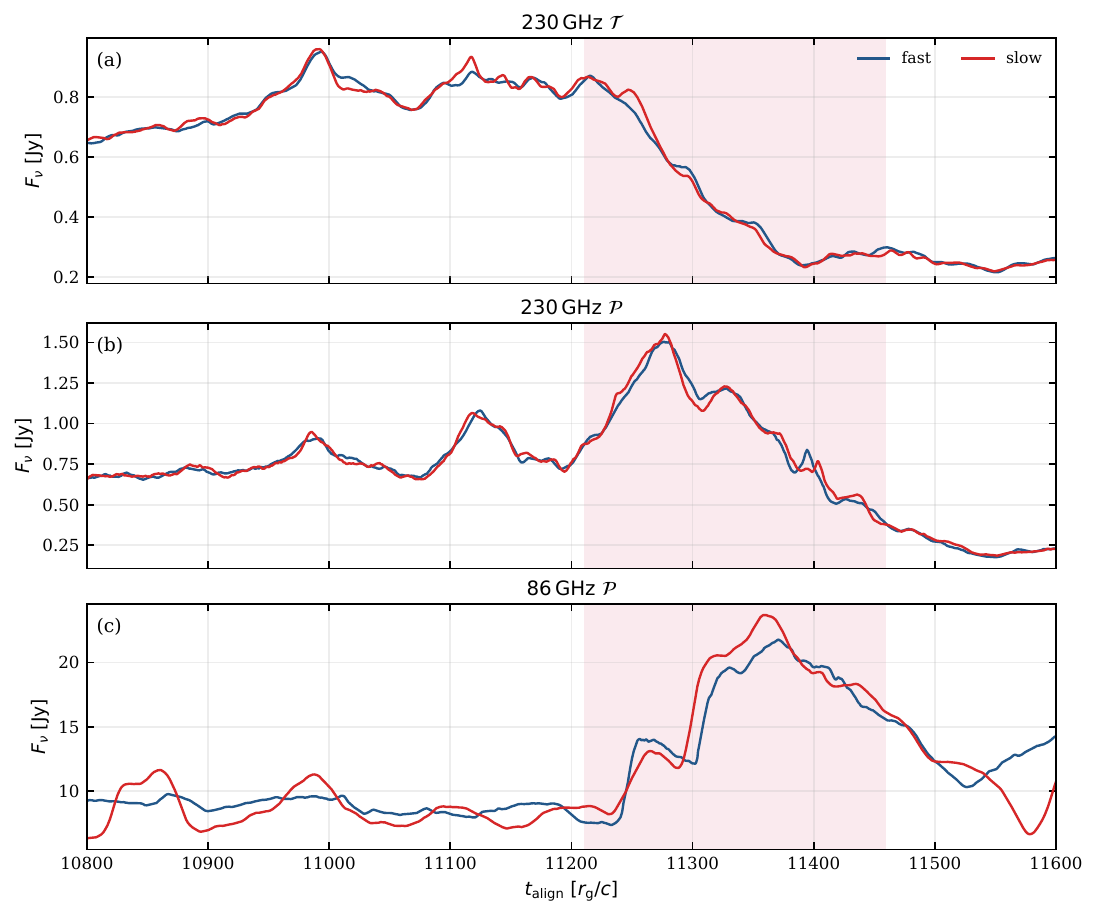}
\caption{Time-aligned fast- and slow-light total fluxes in the three production cases. The first two panels show near-horizon models $\Mc T$ and $\Mc P$, and the third shows jet-image model $\Mc P$, at observing frequencies of $230\,{\rm GHz}$, $230\,{\rm GHz}$, and $86\,{\rm GHz}$, respectively. Pink shading marks the third magnetic-flux eruption. The near-horizon curves are highly correlated after a single global shift. The jet-image curves follow similar overall trends, but their local peaks, amplitudes, and evolutionary details are not fully consistent, as expected when different parts of the extended jet have different path-dependent propagation delays.}
\label{fig:flux-comparison}
\end{figure*}

The near-horizon radiation is concentrated primarily in a compact region, and Figure~\ref{fig:delay-analysis} gives a $\mathrm{p99}$ delay-window width of $110.4\,\rg/c$. The limited delay differences among the dominant emission locations allow the total flux to be approximated by one global time shift. The cross-correlation peaks at $\Delta t_{\rm align}=-31.0\,\rg/c$ and $-28.5\,\rg/c$ for models $\Mc T$ and $\Mc P$, with correlation coefficients of $0.9979$ and $0.9956$, respectively. After shifting, the mean slow-light fluxes exceed the fast-light values by only $0.24\%$ and $0.34\%$, and Equation~\eqref{eq:flux-rms} gives $\delta_{F,{\rm rms}}=2.82\%$ and $4.31\%$. Together, these results show that the global shift captures the principal effect of finite light-travel time on the near-horizon total flux.

For the jet image, a single global time shift provides only partial alignment. It contains radiation from the core, inner jet, and outer jet, and its $\mathrm{p99}$ delay window has a width of $247.1\,\rg/c$, more than twice the near-horizon value. Different jet scales therefore correspond to different fluid times within the same slow-light image. For a relativistic jet moving nearly along the line of sight, the emitting structure continues to evolve outward over these position-dependent propagation delays. A slow-light image is consequently closer to a spatial combination of different fast-light times than to a globally delayed version of any single fast-light image \citep{Tsunetoe:2025slowlight}. Emission from the core and downstream jet enters the same total flux at different evolutionary stages, so a single $\Delta t_{\rm align}$ cannot align all spatial scales simultaneously.

With the expanded search range, the jet total-flux cross-correlation reaches an internal maximum at $\Delta t_{\rm align}=+151.0\,\rg/c$, with a correlation coefficient of $0.9212$. This correlation remains below those of the two near-horizon cases but shows that a single global time shift captures the principal trend shared by the two curves. Over the common comparison interval, the mean slow-light flux is $1.47\%$ lower than the fast-light value, while $\delta_{F,{\rm rms}}$ remains $15.7\%$; Figure~\ref{fig:flux-comparison} also shows that the local peaks, amplitudes, and evolutionary details of the two curves are not fully consistent. A single global time shift therefore recovers only part of the integrated jet variability. The value $+151.0\,\rg/c$ is the best statistical alignment offset of the total-flux curves, not a single physical propagation lag of the jet.

\subsection{Full-Stokes Spatial Differences}\label{subsec:spatial-difference}

\begin{figure*}[t]
\centering
\includegraphics[width=0.94\textwidth]{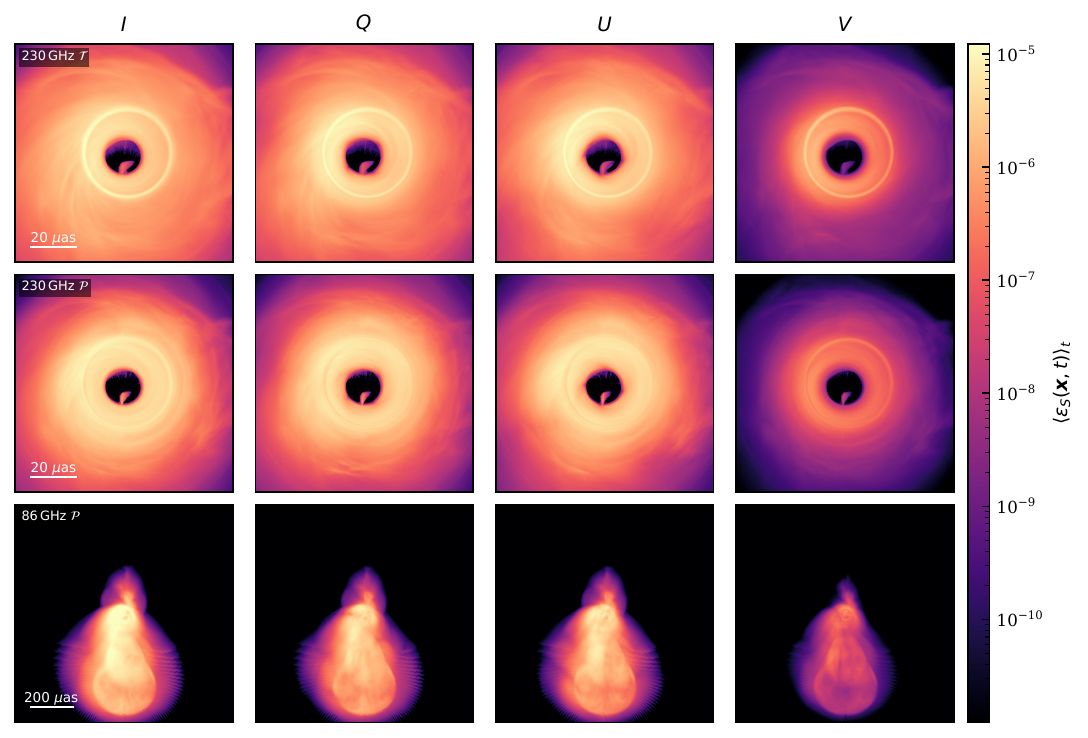}
\caption{Time-aligned full-Stokes spatial differences $\langle\epsilon_S(\boldsymbol{x},t_{\rm align})\rangle_{t_{\rm align}}$ between fast and slow light, as defined in Equation~\eqref{eq:spatial-error}, with normalization by the total Stokes $I$ of the slow-light image at the same time. The three rows show near-horizon model $\Mc T$, near-horizon model $\Mc P$, and jet-image model $\Mc P$; the four columns show Stokes $I/Q/U/V$. All panels share the same logarithmic color scale. The white scale bars are $20\,\mu{\rm as}$ in the first two rows and $200\,\mu{\rm as}$ in the third. Near-horizon differences follow the bright ring and surrounding accretion flow, whereas the jet-image differences extend to jet scales, showing that the spatial morphology of finite-propagation-time effects changes with the size of the emission region.}
\label{fig:stokes-difference}
\end{figure*}

The total flux compared in Section~\ref{subsec:time-alignment} is a spatial integral over the full image, so positive and negative intensity changes in different pixels can cancel. The pixelwise absolute difference retains this spatial information and tests how much image structure remains unrecoverable by the fast-light approximation after the global time shift. We use the slow-light image as the reference and denote the fast- and slow-light Stokes $S$ images at the same alignment time by $S_{\rm fast}$ and $S_{\rm slow}$; the jet image uses the best global time shift of $+151.0\,\rg/c$ given by the internal cross-correlation peak. In Equation~\eqref{eq:spatial-error}, we set $S=S_{\rm fast}$, $S_{\rm ref}=S_{\rm slow}$, and $I_{\rm ref}=I_{\rm slow}$, and average over 1601 aligned times from $t_{\rm align}=10800$ to $11600\,\tg$. By Equation~\eqref{eq:stokes-error}, the sum of the pixelwise contributions equals the full-image difference at the same time, so Figure~\ref{fig:stokes-difference} shows both the magnitude and spatial origin of the difference.

In the near-horizon images, the time-averaged Stokes $I$ spatial differences are $0.270$ and $0.312$ for models $\Mc T$ and $\Mc P$, much larger than the total-flux rms residuals of $2.82\%$ and $4.31\%$ from Equation~\eqref{eq:flux-rms}. The two sets of values are consistent: the flux residual sums the signed pixel differences before taking the residual, whereas the spatial difference takes the absolute value pixel by pixel before summation. The first two rows of Figure~\ref{fig:stokes-difference} further show that the differences are concentrated along the bright ring, the boundary of the central dim region, and the nonaxisymmetric accretion flow outside the ring. These regions both contribute strong emission and contain steep intensity gradients and time-dependent nonaxisymmetric structure. Small differences in the fluid times sampled by different rays therefore change the positions and morphologies of local structures, while most positive and negative intensity changes cancel in the total flux. Near-horizon fast- and slow-light sequences can thus have similar light curves after a global shift while retaining substantial spatial differences.

The linear-polarization structure likewise retains substantial fast/slow-light differences. The Stokes $Q/U$ differences are both $0.266$ for model $\Mc T$ and are $0.244$ and $0.247$ for model $\Mc P$, all of the same order as their respective Stokes $I$ differences. Because $Q$ and $U$ jointly encode the strength and direction of linear polarization, a global shift can align the total flux but cannot synchronize the evolving polarization structure at every image location. The two electron models have similar difference morphologies because they share the same GRMHD dynamics and ray geometry; the electron model primarily changes the radiative weight of each spatial region and hence the difference amplitude. The corresponding Stokes $V$ differences are $0.0249$ and $0.00868$. All four components are normalized here by the total Stokes $I$ of the slow-light image. The smaller $V$ values indicate only that the circular-polarization difference is a smaller fraction of the reference total intensity; they do not establish that the relative variation of circular polarization itself is small.

For model $\Mc P$ in the jet image, the time-averaged $I/Q/U/V$ spatial differences are $0.483$, $0.201$, $0.210$, and $0.00564$, respectively. The Stokes $I$ difference is nearly half of the reference total intensity. Together with the $15.7\%$ flux rms residual found in Section~\ref{subsec:time-alignment}, this shows that the extended emission region retains substantial slow-light differences after the best global time shift. In the third row of Figure~\ref{fig:stokes-difference}, the differences extend continuously from the core into the outer jet, showing that position-dependent propagation delays not only change local intensity but also place different jet scales at different stages of fluid evolution. The extension of the $Q/U$ differences along the jet shows that radiation from different times also mixes in the spatially resolved linear-polarization structure. The present result shows that this mixing affects the overall linear-polarization structure but does not separately isolate the contributions of intrinsic polarization evolution and Faraday propagation.

Current EHT observations remain limited in their constraints on fine spatial structure in continuous near-horizon movies because of angular resolution, dynamic range, and temporal coverage. Integrated quantities such as total flux and integrated linear-polarization fraction are measured more directly \citep{M87_7,Johnson:2023ngEHT}. Under present observing conditions, the near-horizon slow-light effect of similar integrated variability but different spatial structure may therefore not first appear as a large discrepancy in an integrated observable. The next-generation Event Horizon Telescope (ngEHT) will improve angular resolution, dynamic range, and temporal coverage, extending observational targets to finer dynamical structures and spatially resolved polarization images \citep{Johnson:2023ngEHT}. In that regime, local structural differences caused by slow light and evolutionary misalignment among spatial scales can no longer be absorbed by a global time shift. Neglecting finite light-travel time may then cause propagation-induced spatial differences to be attributed incorrectly to accretion-flow dynamics or magnetic-field structure.

\FloatBarrier

\section{Summary and Discussion}\label{sec:summary}

We have presented CoportSL, a contribution-constrained hybrid slow-light framework for time-dependent full-Stokes GRMHD imaging. The method uses the path-weighted contributions of the emission, absorption, Faraday rotation, and Faraday conversion coefficients to identify the slow-light region in which fluid evolution must be retained, and then constructs the effective time window that a single image must access from the propagation-delay distribution within that region. Outside the selected region, the fluid time is fixed but radiative transfer is still evaluated along the complete ray. With a fixed background and grid, the geodesics, parallel-transported polarization bases, and spatial-interpolation information can also be reused throughout the image sequence, concentrating the computational resources on the spatial and temporal ranges that affect time-dependent imaging.

The M87*-like benchmarks show that CoportSL substantially reduces the data requirements of slow light while quantifying the image errors. For the production snapshot interval $\Delta t=0.5\,\tg$, the maximum time-averaged image-interpolation difference relative to the $0.1\,\tg$ reference sequence is $1.88\times10^{-2}$. The maximum difference introduced by the spatial selection relative to the $\mathrm{Full}$ region is $3.98\times10^{-3}$, and that introduced by the $\mathrm{p99}$ window relative to $W_{\rm all}$ is $1.34\times10^{-3}$. Under these independent error measures, the required snapshot-layer counts for the near-horizon and jet images are reduced by $75.3\%$ and $44.7\%$, respectively, and the capacities of the principal data structures are reduced by $69.2\%$ and $41.9\%$. At low core counts, the production hybrid slow-light per-frame radiative-transfer time is comparable to that of fast light.

The fast/slow-light comparison further reveals how delay differences among important radiative locations affect observables. The total-flux correlation coefficients exceed $0.995$ for both near-horizon electron models, yet the time-averaged Stokes $I/Q/U$ spatial differences remain $0.244$--$0.312$; under the same total-intensity normalization, the Stokes $V$ differences are $0.00868$--$0.0249$. Near-horizon images can therefore exhibit similar integrated variability but different spatial structure. For the jet image, the maximum total-flux correlation coefficient is $0.9212$, and the flux rms residual is $15.7\%$. The corresponding Stokes $I/Q/U$ spatial differences are $0.483$, $0.201$, and $0.210$, extending from the core into the outer jet. A global time shift captures the principal trend of the integrated jet variability, but only partially aligns local peaks and amplitudes, while substantial spatial differences remain. The applicability of fast light therefore cannot be judged from integrated variability alone; it depends on whether the delay differences among the important radiative locations can be approximated by a single global time shift.

These quantitative results are based on a single M87*-like MAD sequence on a fixed Kerr background and static grid, together with the viewing direction, electron models, and observing frequencies adopted here. The production snapshot interval, $\mathrm{r50}$, $\Omega_4$, $\mathrm{p99}$, and contribution tolerances are all determined for the present cases. A new source model or target observable requires renewed tests of primitive-variable temporal interpolation, regional transfer-coefficient contributions, discrete-delay coverage, and convergence of the target observable. The present implementation uses the fixed background and grid to reuse the geometric and spatial-interpolation information. A dynamical spacetime or grid would require these data to be updated in time, and the associated accuracy and resource requirements remain to be assessed separately.

Two classes of questions remain unresolved. The first is whether the fast/slow-light differences found here can be identified across a broader set of source models and under realistic interferometric observing conditions. The second is whether fixed candidate partitions, effective windows, and the present parallel implementation can accommodate more complex emission structures and longer time sequences. Addressing the first requires extending the comparison to different accretion states, viewing directions, electron models, and observing frequencies and converting the fast- and slow-light images into time-dependent visibilities and closure quantities with realistic baseline coverage and noise. These tests can determine whether ngEHT dynamical imaging and spatially resolved polarimetry can identify the structural differences caused by slow light. Photon-ring echoes and short-timescale variability statistics also require convergence tests repeated for the corresponding observable to determine the snapshot interval, spatial region, and window coverage. Addressing the second requires adapting the candidate partition and effective window to the target observable and evolving radiative structure, identifying the cause of parallel-scaling saturation at high core counts, and then optimizing data access and parallel computation for long sequences. The present results show that constraining the slow-light calculation with radiative contributions and propagation delays can preserve finite-light-travel-time effects while quantifying the image errors and substantially reducing the data requirements of time-dependent full-Stokes imaging.

\begin{acknowledgments}
The work is partly supported by NSFC Grant No. 12575048, 12275004, and 12588101. This work is also supported by the high-performance computing platform of Peking University. M.G. also acknowledges support from the BNU Tang Scholar Program. OpenAI's ChatGPT \citep{OpenAIChatGPT} was used for language editing, formatting assistance, and consistency checks; the authors reviewed all AI-assisted changes and take full responsibility for the final manuscript.
\end{acknowledgments}

\software{BHAC \citep{Porth2017}, Coport-2.0 \citep{Huang2024, Zhou_2026}, CoportSL, ipole \citep{Moscibrodzka:2017lcu,Moscibrodzka:2021}, NumPy \citep{Harris2020NumPy}, Matplotlib \citep{Hunter2007Matplotlib}}

\section*{Code and Data Availability}

The version of CoportSL used in this work will be made publicly available upon publication at \url{https://github.com/Qfun001/CoportSL} and archived in a persistent repository with a version-specific DOI. Before public release, the manuscript-associated version can be made available to the editors and referees upon request to the authors. The GRMHD snapshots used in this work are not publicly archived because of their large volume but are available from the authors upon reasonable request.

\appendix

\section{Electron Models Used in This Work}\label{app:electron-models}

This Appendix gives the precise definitions of the two electron models used in Section~\ref{subsec:grmhd}. The single-fluid GRMHD simulation provides the total gas internal energy, and the electron temperature is determined using the empirical $R$--$\beta$ relation \citep{Monika2016}:
\begin{equation}
\frac{T_{\rm p}}{T_{\rm e}}
=R_{\rm h}\frac{\beta^2}{1+\beta^2}
+R_{\rm l}\frac{1}{1+\beta^2},
\qquad
\beta=\frac{p_{\rm g}}{p_{\rm b}},
\label{eq:Rbeta}
\end{equation}
where $\rho k_{\rm B}T_{\rm p}=m_{\rm p}p_{\rm g}$, and $p_{\rm g}$ and $p_{\rm b}$ are the gas and magnetic pressures. We take $R_{\rm l}=10$ and $R_{\rm h}=100$ and define the dimensionless electron temperature $\Theta_{\rm e}=k_{\rm B}T_{\rm e}/m_{\rm e}c^2$. The purely thermal model $\Mc T$ uses an isotropic Maxwell--J\"uttner distribution \citep{juttner1911maxwellsche}, with electron number density $n_{\rm e}=\rho/m_{\rm p}$ set by charge neutrality.

Model $\Mc P$ adds an isotropic, nonthermal power-law high-energy tail to the thermal electron background. Its normalized energy distribution is
\begin{equation}
F_p(\gamma)
=\frac{p-1}{\gamma_{\rm min}^{1-p}-\gamma_{\rm max}^{1-p}}
\gamma^{-p},
\qquad
\gamma_{\rm min}\leq\gamma\leq\gamma_{\rm max},
\label{eq:power-law}
\end{equation}
where
\begin{equation}
\gamma_{\rm min}=1+f(\Theta_{\rm e})\Theta_{\rm e},
\qquad
f(\Theta_{\rm e})=\frac{6+15\Theta_{\rm e}}{4+5\Theta_{\rm e}},
\end{equation}
and we take $\gamma_{\rm max}/\gamma_{\rm min}=10^5$ \citep{gammie1998advection,Melzani2014}. The spectral index follows the empirical relation from PIC simulations of magnetic reconnection \citep{Ball_2018},
\begin{equation}
p(\beta,\sigma_{\rm M})
=1.8+0.7\sigma_{\rm M}^{-1/2}
+3.7\sigma_{\rm M}^{-0.19}
\tanh\left(23.4\sigma_{\rm M}^{0.26}\beta\right),
\label{eq:power-index}
\end{equation}
with the restriction $2.001\leq p\leq10$.

The local nonthermal-electron energy fraction is \citep{Ball_2018}
\begin{equation}
\begin{aligned}
\epsilon(\beta,\sigma_{\rm M})
&=1-\left(4.2\sigma_{\rm M}^{0.55}+1\right)^{-1}\\
&\quad+0.64\sigma_{\rm M}^{0.07}
\tanh\left(-68\sigma_{\rm M}^{0.13}\beta\right),
\end{aligned}
\label{eq:epsilon}
\end{equation}
with the restriction $0\leq\epsilon\leq1$. The mean dimensionless energies of the thermal and nonthermal electrons are
\begin{equation}
u_{\rm th}=f(\Theta_{\rm e})\Theta_{\rm e},
\qquad
u_{\rm nth}=\frac{p-1}{p-2}\gamma_{\rm min}-1,
\end{equation}
where $u_{\rm nth}$ uses the approximation $\gamma_{\rm max}\gg\gamma_{\rm min}$ for $p>2$. The nonthermal electron number fraction is
\begin{equation}
f_{\rm nth}
=\frac{\epsilon u_{\rm th}}
{(1-\epsilon)u_{\rm nth}+\epsilon u_{\rm th}}.
\label{eq:fnth}
\end{equation}
The local emission, absorption, and Faraday coefficients are combined as
\begin{equation}
c_\nu=(1-f_{\rm nth})c_{{\rm th},\nu}
+f_{\rm nth}c_{{\rm nth},\nu}
\label{eq:coeff-mix}
\end{equation}
where $c_\nu$ denotes $j_\nu$, $\alpha_\nu$, or $\rho_\nu$. The thermal and power-law electron coefficients use the analytic fitting formulae of \citet{Dexter:2016cdk,Marszewski:2021fkr}.

\clearpage
\bibliographystyle{aasjournalv7}
\bibliography{main}

@article{Porth2017,
	author = {Porth, Oliver and Olivares, Hector and Mizuno, Yosuke and Younsi, Ziri and Rezzolla, Luciano and Mo{\'s}cibrodzka, Monika and Falcke, Heino and Kramer, Michael},
	title = {The black hole accretion code},
	journal = {Computational Astrophysics and Cosmology},
	year = {2017},
	volume = {4},
	number = {1},
	pages = {1},
	issn = {2197-7909},
	doi = {10.1186/s40668-017-0020-2},
	url = {https://doi.org/10.1186/s40668-017-0020-2},
}

@article{Huang2024,
	author = "Huang, Jiewei and Zheng, Liheng and Guo, Minyong and Chen, Bin",
	title = "{Coport: a new public code for polarized radiative transfer in a covariant framework}",
	eprint = "2407.10431",
	archivePrefix = "arXiv",
	primaryClass = "astro-ph.HE",
	doi = "10.1088/1475-7516/2024/11/054",
	journal = "JCAP",
	volume = "2024",
	number = "11",
	pages = "054",
	year = "2024"
}

@article{Monika2016,
	author = {Mo{\'s}cibrodzka, Monika and Falcke, Heino and Shiokawa, Hotaka},
	title = {General relativistic magnetohydrodynamical simulations of the jet in {M87}},
	DOI= "10.1051/0004-6361/201526630",
	url= "https://doi.org/10.1051/0004-6361/201526630",
	journal = {A\&A},
	year = 2016,
	volume = 586,
	pages = "A38",
	month = "",
}

@article{M87_1,
	author = "Akiyama, Kazunori and Alberdi, Antxon and Alef, Walter and Asada, Keiichi and Azulay, Rebecca and Baczko, Anne-Kathrin",
	collaboration = "Event Horizon Telescope",
	title = "{First M87 Event Horizon Telescope Results. I. The Shadow of the Supermassive Black Hole}",
	eprint = "1906.11238",
	archivePrefix = "arXiv",
	primaryClass = "astro-ph.GA",
	doi = "10.3847/2041-8213/ab0ec7",
	journal = "Astrophys. J. Lett.",
	volume = "875",
	pages = "L1",
	year = "2019"
}

@article{M87_4,
	author = "Akiyama, Kazunori and Alberdi, Antxon and Alef, Walter and Asada, Keiichi and Azulay, Rebecca and Baczko, Anne-Kathrin",
	collaboration = "Event Horizon Telescope",
	title = "{First M87 Event Horizon Telescope Results. IV. Imaging the Central Supermassive Black Hole}",
	eprint = "1906.11241",
	archivePrefix = "arXiv",
	primaryClass = "astro-ph.GA",
	doi = "10.3847/2041-8213/ab0e85",
	journal = "Astrophys. J. Lett.",
	volume = "875",
	number = "1",
	pages = "L4",
	year = "2019"
}

@article{M87_5,
	author = "Akiyama, Kazunori and Alberdi, Antxon and Alef, Walter and Asada, Keiichi and Azulay, Rebecca and Baczko, Anne-Kathrin",
	collaboration = "Event Horizon Telescope",
	title = "{First M87 Event Horizon Telescope Results. V. Physical Origin of the Asymmetric Ring}",
	eprint = "1906.11242",
	archivePrefix = "arXiv",
	primaryClass = "astro-ph.GA",
	doi = "10.3847/2041-8213/ab0f43",
	journal = "Astrophys. J. Lett.",
	volume = "875",
	number = "1",
	pages = "L5",
	year = "2019"
}

@article{M87_7,
	author = "Akiyama, Kazunori and Algaba, Juan Carlos and Alberdi, Antxon and Alef, Walter and Anantua, Richard and Asada, Keiichi",
	collaboration = "Event Horizon Telescope",
	title = "{First M87 Event Horizon Telescope Results. VII. Polarization of the Ring}",
	eprint = "2105.01169",
	archivePrefix = "arXiv",
	primaryClass = "astro-ph.HE",
	reportNumber = "FERMILAB-PUB-21-849-PPD",
	doi = "10.3847/2041-8213/abe71d",
	journal = "Astrophys. J. Lett.",
	volume = "910",
	number = "1",
	pages = "L12",
	year = "2021"
}

@article{M87_8,
	author = "Akiyama, Kazunori and Algaba, Juan Carlos and Alberdi, Antxon and Alef, Walter and Anantua, Richard and Asada, Keiichi",
	collaboration = "Event Horizon Telescope",
	title = "{First M87 Event Horizon Telescope Results. VIII. Magnetic Field Structure near The Event Horizon}",
	eprint = "2105.01173",
	archivePrefix = "arXiv",
	primaryClass = "astro-ph.HE",
	reportNumber = "FERMILAB-PUB-21-850-PPD",
	doi = "10.3847/2041-8213/abe4de",
	journal = "Astrophys. J. Lett.",
	volume = "910",
	number = "1",
	pages = "L13",
	year = "2021"
}

@article{M87_9,
	author = "Akiyama, Kazunori and Alberdi, Antxon and Alef, Walter and Algaba, Juan Carlos and Anantua, Richard and Asada, Keiichi",
	collaboration = "Event Horizon Telescope",
	title = "{First M87 Event Horizon Telescope Results. IX. Detection of Near-horizon Circular Polarization}",
	eprint = "2311.10976",
	archivePrefix = "arXiv",
	primaryClass = "astro-ph.HE",
	doi = "10.3847/2041-8213/acff70",
	journal = "Astrophys. J. Lett.",
	volume = "957",
	number = "2",
	pages = "L20",
	year = "2023"
}

@article{M87_2018_I,
	author = "{Event Horizon Telescope Collaboration}",
	collaboration = "Event Horizon Telescope",
	title = "{The persistent shadow of the supermassive black hole of M 87. I. Observations, calibration, imaging, and analysis}",
	doi = "10.1051/0004-6361/202347932",
	journal = "Astron. Astrophys.",
	volume = "681",
	pages = "A79",
	year = "2024"
}

@article{M87_2018_II,
	author = "{Event Horizon Telescope Collaboration}",
	collaboration = "Event Horizon Telescope",
	title = "{The persistent shadow of the supermassive black hole of M87. II. Model comparisons and theoretical interpretations}",
	doi = "10.1051/0004-6361/202451296",
	journal = "Astron. Astrophys.",
	volume = "693",
	pages = "A265",
	year = "2025"
}

@article{SgrA_1,
	author = "Akiyama, Kazunori and Alberdi, Antxon and Alef, Walter and Algaba, Juan Carlos and Anantua, Richard and Asada, Keiichi",
	collaboration = "Event Horizon Telescope",
	title = "{First Sagittarius A* Event Horizon Telescope Results. I. The Shadow of the Supermassive Black Hole in the Center of the Milky Way}",
	eprint = "2311.08680",
	archivePrefix = "arXiv",
	primaryClass = "astro-ph.HE",
	doi = "10.3847/2041-8213/ac6674",
	journal = "Astrophys. J. Lett.",
	volume = "930",
	number = "2",
	pages = "L12",
	year = "2022"
}

@article{SgrA_3,
	author = "Akiyama, Kazunori and Alberdi, Antxon and Alef, Walter and Algaba, Juan Carlos and Anantua, Richard and Asada, Keiichi",
	collaboration = "Event Horizon Telescope",
	title = "{First Sagittarius A* Event Horizon Telescope Results. III. Imaging of the Galactic Center Supermassive Black Hole}",
	eprint = "2311.09479",
	archivePrefix = "arXiv",
	primaryClass = "astro-ph.HE",
	doi = "10.3847/2041-8213/ac6429",
	journal = "Astrophys. J. Lett.",
	volume = "930",
	number = "2",
	pages = "L14",
	year = "2022"
}

@article{SgrA_4,
	author = "Akiyama, Kazunori and Alberdi, Antxon and Alef, Walter and Algaba, Juan Carlos and Anantua, Richard and Asada, Keiichi",
	collaboration = "Event Horizon Telescope",
	title = "{First Sagittarius A* Event Horizon Telescope Results. IV. Variability, Morphology, and Black Hole Mass}",
	eprint = "2311.08697",
	archivePrefix = "arXiv",
	primaryClass = "astro-ph.HE",
	reportNumber = "FERMILAB-PUB-22-423-PPD",
	doi = "10.3847/2041-8213/ac6736",
	journal = "Astrophys. J. Lett.",
	volume = "930",
	number = "2",
	pages = "L15",
	year = "2022"
}

@article{SgrA_5,
	author = "Akiyama, Kazunori and Alberdi, Antxon and Alef, Walter and Algaba, Juan Carlos and Anantua, Richard and Asada, Keiichi",
	collaboration = "Event Horizon Telescope",
	title = "{First Sagittarius A* Event Horizon Telescope Results. V. Testing Astrophysical Models of the Galactic Center Black Hole}",
	eprint = "2311.09478",
	archivePrefix = "arXiv",
	primaryClass = "astro-ph.HE",
	reportNumber = "FERMILAB-PUB-22-419-PPD",
	doi = "10.3847/2041-8213/ac6672",
	journal = "Astrophys. J. Lett.",
	volume = "930",
	number = "2",
	pages = "L16",
	year = "2022"
}

@article{SgrA_6,
	author = "Akiyama, Kazunori and Alberdi, Antxon and Alef, Walter and Algaba, Juan Carlos and Anantua, Richard and Asada, Keiichi",
	collaboration = "Event Horizon Telescope",
	title = "{First Sagittarius A* Event Horizon Telescope Results. VI. Testing the Black Hole Metric}",
	eprint = "2311.09484",
	archivePrefix = "arXiv",
	primaryClass = "astro-ph.HE",
	reportNumber = "FERMILAB-PUB-22-422-PPD",
	doi = "10.3847/2041-8213/ac6756",
	journal = "Astrophys. J. Lett.",
	volume = "930",
	number = "2",
	pages = "L17",
	year = "2022"
}

@article{SgrA_7,
	author = "Akiyama, Kazunori and Alberdi, Antxon and Alef, Walter and Algaba, Juan Carlos and Anantua, Richard and Asada, Keiichi",
	collaboration = "Event Horizon Telescope",
	title = "{First Sagittarius A* Event Horizon Telescope Results. VII. Polarization of the Ring}",
	doi = "10.3847/2041-8213/ad2df0",
	journal = "Astrophys. J. Lett.",
	volume = "964",
	number = "2",
	pages = "L25",
	year = "2024"
}

@article{SgrA_8,
	author = "Akiyama, Kazunori and Alberdi, Antxon and Alef, Walter and Algaba, Juan Carlos and Anantua, Richard and Asada, Keiichi",
	collaboration = "Event Horizon Telescope",
	title = "{First Sagittarius A* Event Horizon Telescope Results. VIII. Physical Interpretation of the Polarized Ring}",
	doi = "10.3847/2041-8213/ad2df1",
	journal = "Astrophys. J. Lett.",
	volume = "964",
	number = "2",
	pages = "L26",
	year = "2024"
}

@article{Ball_2018,
	doi = {10.3847/1538-4357/aac820},
	url = {https://dx.doi.org/10.3847/1538-4357/aac820},
	year = {2018},
	month = {jul},
	publisher = {The American Astronomical Society},
	volume = {862},
	number = {1},
	pages = {80},
	author = {Ball, David and Sironi, Lorenzo and {\"O}zel, Feryal},
	title = {{Electron and Proton Acceleration in Trans-relativistic Magnetic Reconnection: Dependence on Plasma Beta and Magnetization}},
	journal = {The Astrophysical Journal},
}

@article{Melzani2014,
	author = {Melzani, Micka{\"e}l and Walder, Rolf and Folini, Doris and Winisdoerffer, Christophe and Favre, Jean M.},
	title = {The energetics of relativistic magnetic reconnection: ion-electron repartition and particle distribution hardness},
	DOI= "10.1051/0004-6361/201424193",
	url= "https://doi.org/10.1051/0004-6361/201424193",
	journal = {A\&A},
	year = 2014,
	volume = 570,
	pages = "A112",
	month = "",
}

@article{Narayan2003,
	author = {Narayan, Ramesh and Igumenshchev, Igor V. and Abramowicz, Marek A.},
	title = {Magnetically Arrested Disk: an Energetically Efficient Accretion Flow},
	journal = {Publications of the Astronomical Society of Japan},
	volume = {55},
	number = {6},
	pages = {L69--L72},
	year = {2003},
	month = {12},
	issn = {0004-6264},
	doi = {10.1093/pasj/55.6.L69},
	url = {https://doi.org/10.1093/pasj/55.6.L69},
}

@article{McKinney2004,
	doi = {10.1086/422244},
	url = {https://dx.doi.org/10.1086/422244},
	year = {2004},
	month = {aug},
	publisher = {},
	volume = {611},
	number = {2},
	pages = {977--995},
	author = {McKinney, Jonathan C. and Gammie, Charles F.},
	title = {A Measurement of the Electromagnetic Luminosity of a Kerr Black Hole},
	journal = {The Astrophysical Journal},
}

@article{Gammie_2003,
	doi = {10.1086/374594},
	url = {https://dx.doi.org/10.1086/374594},
	year = {2003},
	month = {may},
	publisher = {},
	volume = {589},
	number = {1},
	pages = {444--457},
	author = {Gammie, Charles F. and McKinney, Jonathan C. and T{\'o}th, G{\'a}bor},
	title = {HARM: A Numerical Scheme for General Relativistic Magnetohydrodynamics},
	journal = {The Astrophysical Journal},
}

@article{gammie1998advection,
	title={Advection-dominated accretion flows in the Kerr metric. I. Basic equations},
	author={Gammie, Charles F and Popham, Robert},
	journal={The Astrophysical Journal},
	volume={498},
	number={1},
	pages={313--326},
	year={1998},
	doi={10.1086/305521},
	url={https://doi.org/10.1086/305521},
	publisher={IOP Publishing}
}

@article{Mizuno:2021esc,
	author = "Mizuno, Yosuke and Fromm, Christian M. and Younsi, Ziri and Porth, Oliver and Olivares, Hector and Rezzolla, Luciano",
	title = "{Comparison of the ion-to-electron temperature ratio prescription: GRMHD simulations with electron thermodynamics}",
	eprint = "2106.09272",
	archivePrefix = "arXiv",
	primaryClass = "astro-ph.HE",
	doi = "10.1093/mnras/stab1753",
	journal = "Mon. Not. Roy. Astron. Soc.",
	volume = "506",
	number = "1",
	pages = "741--758",
	year = "2021"
}

@article{Davelaar:2019jxr,
	author = "Davelaar, J. and Olivares, H. and Porth, O. and Bronzwaer, T. and Janssen, M. and Roelofs, F. and Mizuno, Y. and Fromm, C. M. and Falcke, H. and Rezzolla, L.",
	title = "{Modeling non-thermal emission from the jet-launching region of M 87 with adaptive mesh refinement}",
	eprint = "1906.10065",
	archivePrefix = "arXiv",
	primaryClass = "astro-ph.HE",
	doi = "10.1051/0004-6361/201936150",
	journal = "Astron. Astrophys.",
	volume = "632",
	pages = "A2",
	year = "2019"
}

@article{Zhang:2024ddt,
	author = "Zhang, Mingyuan and Mizuno, Yosuke and Fromm, Christian M. and Younsi, Ziri and Cruz-Osorio, Alejandro",
	title = "{Impacts of nonthermal emission on the images of a black hole shadow and extended jets in two-temperature GRMHD simulations}",
	eprint = "2404.04033",
	archivePrefix = "arXiv",
	primaryClass = "astro-ph.HE",
	doi = "10.1051/0004-6361/202449497",
	journal = "Astron. Astrophys.",
	volume = "687",
	pages = "A88",
	year = "2024"
}

@article{Tchekhovskoy:2011zx,
	author = "Tchekhovskoy, Alexander and Narayan, Ramesh and McKinney, Jonathan C.",
	title = "{Efficient Generation of Jets from Magnetically Arrested Accretion on a Rapidly Spinning Black Hole}",
	eprint = "1108.0412",
	archivePrefix = "arXiv",
	primaryClass = "astro-ph.HE",
	doi = "10.1111/j.1745-3933.2011.01147.x",
	journal = "Mon. Not. Roy. Astron. Soc.",
	volume = "418",
	pages = "L79--L83",
	year = "2011"
}

@article{Fromm:2021mqd,
	author = "Fromm, Christian M. and Cruz-Osorio, Alejandro and Mizuno, Yosuke and Nathanail, Antonios and Younsi, Ziri and Porth, Oliver and Olivares, Hector and Davelaar, Jordy and Falcke, Heino and Kramer, Michael and Rezzolla, Luciano",
	title = "{Impact of non-thermal particles on the spectral and structural properties of M87}",
	eprint = "2111.02518",
	archivePrefix = "arXiv",
	primaryClass = "astro-ph.HE",
	doi = "10.1051/0004-6361/202142295",
	journal = "Astron. Astrophys.",
	volume = "660",
	pages = "A107",
	year = "2022"
}

@article{juttner1911maxwellsche,
	title={Das maxwellsche gesetz der geschwindigkeitsverteilung in der relativtheorie},
	author={J{\"u}ttner, Ferencz},
	journal={Annalen der Physik},
	volume={339},
	number={5},
	pages={856--882},
	year={1911},
	doi={10.1002/andp.19113390503},
	url={https://doi.org/10.1002/andp.19113390503},
	publisher={Wiley Online Library}
}

@article{Dexter:2016cdk,
	author = "Dexter, Jason",
	title = "{A public code for general relativistic, polarised radiative transfer around spinning black holes}",
	eprint = "1602.03184",
	archivePrefix = "arXiv",
	primaryClass = "astro-ph.HE",
	doi = "10.1093/mnras/stw1526",
	journal = "Mon. Not. Roy. Astron. Soc.",
	volume = "462",
	number = "1",
	pages = "115--136",
	year = "2016"
}

@article{Marszewski:2021fkr,
	author = "Marszewski, Andrew and Prather, Ben S. and Joshi, Abhishek V. and Pandya, Alex and Gammie, Charles F.",
	title = "{Updated Transfer Coefficients for Magnetized Plasmas}",
	eprint = "2108.10359",
	archivePrefix = "arXiv",
	primaryClass = "astro-ph.HE",
	doi = "10.3847/1538-4357/ac1b28",
	journal = "Astrophys. J.",
	volume = "921",
	number = "1",
	pages = "17",
	year = "2021"
}

@article{Broderick:2003fc,
	author = "Broderick, Avery and Blandford, Roger",
	title = "{Covariant magnetoionic theory. 2. Radiative transfer}",
	eprint = "astro-ph/0311360",
	archivePrefix = "arXiv",
	reportNumber = "SLAC-PUB-10262",
	doi = "10.1111/j.1365-2966.2004.07582.x",
	journal = "Mon. Not. Roy. Astron. Soc.",
	volume = "349",
	pages = "994--1008",
	year = "2004"
}

@article{Shcherbakov:2010kh,
	author = "Shcherbakov, Roman V. and Huang, Lei",
	title = "{General relativistic polarized radiative transfer: building a dynamics-observations interface}",
	eprint = "1007.4831",
	archivePrefix = "arXiv",
	primaryClass = "astro-ph.HE",
	doi = "10.1111/j.1365-2966.2010.17502.x",
	journal = "Mon. Not. Roy. Astron. Soc.",
	volume = "410",
	pages = "1052--1063",
	year = "2011"
}

@article{noble2007simulating,
	title={Simulating the emission and outflows from accretion discs},
	author={Noble, Scott C and Leung, Po Kin and Gammie, Charles F and Book, Laura G},
	journal={Classical and Quantum Gravity},
	volume={24},
	number={12},
	pages={S259--S274},
	year={2007},
	doi={10.1088/0264-9381/24/12/S17},
	url={https://doi.org/10.1088/0264-9381/24/12/S17},
	publisher={IOP Publishing}
}

@article{moscibrodzka2009radiative,
	title={Radiative models of Sgr A* from GRMHD simulations},
	author={Mo{\'s}cibrodzka, Monika and Gammie, Charles F and Dolence, Joshua C and Shiokawa, Hotaka and Leung, Po Kin},
	journal={The Astrophysical Journal},
	volume={706},
	number={1},
	pages={497},
	year={2009},
	doi={10.1088/0004-637X/706/1/497},
	url={https://doi.org/10.1088/0004-637X/706/1/497},
	publisher={IOP Publishing}
}

@article{Gold:2016hld,
	author = "Gold, Roman and McKinney, Jonathan C. and Johnson, Michael D. and Doeleman, Sheperd S.",
	title = "{Probing the Magnetic Field Structure in $\mathrm{Sgr}\,{\rm{A}}* $ on Black Hole Horizon Scales with Polarized Radiative Transfer Simulations}",
	eprint = "1601.05550",
	archivePrefix = "arXiv",
	primaryClass = "astro-ph.HE",
	doi = "10.3847/1538-4357/aa6193",
	journal = "Astrophys. J.",
	volume = "837",
	number = "2",
	pages = "180",
	year = "2017"
}

@article{EventHorizonTelescope:2025vum,
	author = "Akiyama, Kazunori and Albentosa-Ruiz, Ezequiel and Alberdi, Antxon and Alef, Walter and Algaba, Juan Carlos and Anantua, Richard",
	collaboration = "Event Horizon Telescope",
	title = "{Horizon-scale variability of M87* from 2017--2021 EHT observations}",
	eprint = "2509.24593",
	archivePrefix = "arXiv",
	primaryClass = "astro-ph.HE",
	doi = "10.1051/0004-6361/202555855",
	journal = "Astron. Astrophys.",
	volume = "704",
	pages = "A91",
	month = "12",
	year = "2025"
}

@article{Tsunetoe:2025slowlight,
	author = "Tsunetoe, Yuh and Pesce, Dominic W. and Narayan, Ramesh",
	title = "{Slow-Light Effect in the Jet-Launching Region of M87}",
	eprint = "2512.02113",
	archivePrefix = "arXiv",
	primaryClass = "astro-ph.HE",
	doi = "10.3847/1538-4357/ae43e7",
	journal = "Astrophys. J.",
	volume = "1000",
	number = "1",
	pages = "29",
	year = "2026"
}

@article{Najafi-Ziyazi:2023oil,
	author = "Najafi-Ziyazi, Mahdi and Davelaar, Jordy and Mizuno, Yosuke and Porth, Oliver",
	title = "{Flares in the Galactic centre {\textendash} II. Polarization signatures of flares at mm-wavelengths}",
	eprint = "2308.16740",
	archivePrefix = "arXiv",
	primaryClass = "astro-ph.HE",
	doi = "10.1093/mnras/stae1343",
	journal = "Mon. Not. Roy. Astron. Soc.",
	volume = "531",
	number = "4",
	pages = "3961--3972",
	year = "2024"
}

@article{Hou:2024qqo,
	author = "Hou, Yehui and Huang, Jiewei and Guo, Minyong and Mizuno, Yosuke and Chen, Bin",
	title = "{Near-horizon Polarization as a Diagnostic of Black Hole Spacetime}",
	eprint = "2409.07248",
	archivePrefix = "arXiv",
	primaryClass = "gr-qc",
	reportNumber = "988 L51",
	doi = "10.3847/2041-8213/adee09",
	journal = "Astrophys. J. Lett.",
	volume = "988",
	number = "2",
	pages = "L51",
	year = "2025"
}

@article{EventHorizonTelescope:2019pcy,
	author = "Porth, Oliver and Chatterjee, Koushik and Narayan, Ramesh and Gammie, Charles F. and Mizuno, Yosuke and Anninos, Peter",
	collaboration = "Event Horizon Telescope",
	title = "{The Event Horizon General Relativistic Magnetohydrodynamic Code Comparison Project}",
	eprint = "1904.04923",
	archivePrefix = "arXiv",
	primaryClass = "astro-ph.HE",
	doi = "10.3847/1538-4365/ab29fd",
	journal = "Astrophys. J. Suppl.",
	volume = "243",
	number = "2",
	pages = "26",
	year = "2019"
}

@article{EventHorizonTelescope:2021dqv,
	author = "Kocherlakota, Prashant and Rezzolla, Luciano and Falcke, Heino and Fromm, Christian M. and Kramer, Michael and Mizuno, Yosuke",
	collaboration = "Event Horizon Telescope",
	title = "{Constraints on black-hole charges with the 2017 EHT observations of M87*}",
	eprint = "2105.09343",
	archivePrefix = "arXiv",
	primaryClass = "gr-qc",
	reportNumber = "FERMILAB-PUB-21-847-PPD",
	doi = "10.1103/PhysRevD.103.104047",
	journal = "Phys. Rev. D",
	volume = "103",
	number = "10",
	pages = "104047",
	year = "2021"
}

@article{EventHorizonTelescope:2020qrl,
	author = "Psaltis, Dimitrios and Medeiros, Lia and Christian, Pierre and {\"O}zel, Feryal and Akiyama, Kazunori and Alberdi, Antxon",
	collaboration = "Event Horizon Telescope",
	title = "{Gravitational Test Beyond the First Post-Newtonian Order with the Shadow of the M87 Black Hole}",
	eprint = "2010.01055",
	archivePrefix = "arXiv",
	primaryClass = "gr-qc",
	doi = "10.1103/PhysRevLett.125.141104",
	journal = "Phys. Rev. Lett.",
	volume = "125",
	number = "14",
	pages = "141104",
	year = "2020"
}

@article{Jiang:2025huk,
	author = "Jiang, Hong-Xuan and Mizuno, Yosuke and Dihingia, Indu K. and Yuan, Feng and Lin, Xi and Fromm, Christian M. and Nathanail, Antonios and Younsi, Ziri",
	title = "{The Physical Origin and Time Lag of Multifrequency Flares from Sgr A*}",
	eprint = "2507.12789",
	archivePrefix = "arXiv",
	primaryClass = "astro-ph.HE",
	doi = "10.3847/1538-4357/adf1e5",
	journal = "Astrophys. J.",
	volume = "990",
	number = "1",
	pages = "81",
	year = "2025"
}

@article{degl1985solution,
  title={On the solution of the radiative transfer equations for polarized radiation},
  author={Degl'Innocenti, E Landi and Degl'Innocenti, M Landi},
  journal={Solar Physics},
  volume={97},
  number={2},
  pages={239--250},
  year={1985},
  doi={10.1007/BF00165988},
  url={https://doi.org/10.1007/BF00165988},
  publisher={Springer}
}

@article{Moscibrodzka:2017lcu,
    author = "Mo{\'s}cibrodzka, Monika and Gammie, Charles F.",
    title = "{ipole {\textendash} semi-analytic scheme for relativistic polarized radiative transport}",
    eprint = "1712.03057",
    archivePrefix = "arXiv",
    primaryClass = "astro-ph.HE",
    doi = "10.1093/mnras/stx3162",
    journal = "Mon. Not. Roy. Astron. Soc.",
    volume = "475",
    number = "1",
    pages = "43--54",
    year = "2018"
}

@article{Bronzwaer:2018,
    author = "Bronzwaer, Thomas and Davelaar, Jordy and Younsi, Ziri and Mo{\'s}cibrodzka, Monika and Falcke, Heino and Kramer, Michael and Rezzolla, Luciano",
    title = "{RAPTOR I: Time-dependent radiative transfer in arbitrary spacetimes}",
    eprint = "1801.10452",
    archivePrefix = "arXiv",
    primaryClass = "astro-ph.HE",
    journal = "Astronomy and Astrophysics",
    volume = "613",
    pages = "A2",
    year = "2018",
    doi = "10.1051/0004-6361/201732149"
}

@article{Dexter2010,
    author = "Dexter, Jason and Agol, Eric and Fragile, P. Chris and McKinney, Jonathan C.",
    title = "{The Submillimeter Bump in Sgr A* from Relativistic MHD Simulations}",
    eprint = "1005.4062",
    archivePrefix = "arXiv",
    primaryClass = "astro-ph.HE",
    journal = "The Astrophysical Journal",
    volume = "717",
    number = "2",
    pages = "1092--1104",
    year = "2010",
    doi = "10.1088/0004-637X/717/2/1092"
}

@article{White:2022,
    author = "White, Christopher J.",
    title = "{Blacklight: A General-relativistic Ray-tracing and Analysis Tool}",
    eprint = "2203.15963",
    archivePrefix = "arXiv",
    primaryClass = "astro-ph.HE",
    journal = "The Astrophysical Journal Supplement Series",
    volume = "262",
    number = "1",
    pages = "28",
    year = "2022",
    doi = "10.3847/1538-4365/ac77ef"
}

@article{Moscibrodzka:2021,
    author = "Mo{\'s}cibrodzka, Monika and Janiuk, Agnieszka and De Laurentis, Maria",
    title = "{Unraveling Circular Polarimetric Images of Magnetically Arrested Accretion Flows near Event Horizon of a Black Hole}",
    eprint = "2103.00267",
    archivePrefix = "arXiv",
    primaryClass = "astro-ph.HE",
    journal = "Monthly Notices of the Royal Astronomical Society",
    volume = "508",
    number = "3",
    pages = "4282--4296",
    year = "2021",
    doi = "10.1093/mnras/stab2790"
}

@article{Vos:2024,
    author = "Vos, Jesse and Davelaar, Jordy and Olivares, Hector and Brinkerink, Christiaan and Falcke, Heino",
    title = "{Magnetic flux eruptions at the root of time lags in low-luminosity AGN}",
    journal = "Astronomy and Astrophysics",
    volume = "689",
    pages = "A112",
    year = "2024",
    doi = "10.1051/0004-6361/202449265"
}

@article{Conroy:2023,
    author = "Conroy, Nicholas S. and Baub{\"o}ck, Michi and Dhruv, Vedant and Lee, De-Hwan and Broderick, Avery E. and Chan, Chi-Kwan and Georgiev, Borislav and Joshi, Abhishek V. and Prather, Ben S. and Gammie, Charles F.",
    title = "{Rotation in Event Horizon Telescope Movies}",
    eprint = "2304.03826",
    archivePrefix = "arXiv",
    primaryClass = "astro-ph.HE",
    journal = "The Astrophysical Journal",
    volume = "951",
    number = "1",
    pages = "46",
    year = "2023",
    doi = "10.3847/1538-4357/acd2c8"
}

@article{Wong:2024Echoes,
    author = "Wong, George N. and Medeiros, Lia and C{\'a}rdenas-Avenda{\~n}o, Alejandro and Stone, James M.",
    title = "{Measuring Black Hole Light Echoes with Very Long Baseline Interferometry}",
    eprint = "2410.10950",
    archivePrefix = "arXiv",
    primaryClass = "astro-ph.HE",
    journal = "The Astrophysical Journal Letters",
    volume = "975",
    number = "2",
    pages = "L40",
    year = "2024",
    doi = "10.3847/2041-8213/ad8650"
}

@article{Hadar:2021,
    author = "Hadar, Shahar and Johnson, Michael D. and Lupsasca, Alexandru and Wong, George N.",
    title = "{Photon Ring Autocorrelations}",
    eprint = "2010.03683",
    archivePrefix = "arXiv",
    primaryClass = "gr-qc",
    journal = "Physical Review D",
    volume = "103",
    number = "10",
    pages = "104038",
    year = "2021",
    doi = "10.1103/PhysRevD.103.104038"
}

@article{RojasPaternina:2026,
    author = "Rojas-Paternina, Daniel and C{\'a}rdenas-Avenda{\~n}o, Alejandro",
    title = "{Light Propagation Prescriptions for Black Hole Movies}",
    eprint = "2605.12659",
    archivePrefix = "arXiv",
    primaryClass = "astro-ph.HE",
    doi = "10.1103/nsg2-hy9p",
    journal = "Phys. Rev. D",
    volume = "114",
    number = "2",
    pages = "023031",
    year = "2026"
}

@article{Bezdekova:2026,
    author = "Bezd{\v e}kov{\'a}, Barbora and Hadar, Shahar and Wong, George N. and Wielgus, Maciek",
    title = "{Extreme-Lensing Signatures Revealed by Correlations of Simulated Black-Hole Movies}",
    eprint = "2512.09641",
    archivePrefix = "arXiv",
    primaryClass = "astro-ph.HE",
    doi = "10.1038/s41550-026-02874-x",
    journal = "Nature Astron.",
    volume = "10",
    number = "8",
    pages = "1199--1207",
    year = "2026"
}

@article{Zhang:2025vyx,
	author = "Zhang, Zhenyu and Hou, Yehui and Guo, Minyong and Mizuno, Yosuke and Chen, Bin",
	title = "{Autocorrelation signatures in time-resolved black hole flare images: Secondary peaks and convergence structure}",
	eprint = "2503.17200",
	archivePrefix = "arXiv",
	primaryClass = "astro-ph.HE",
	doi = "10.1103/zmnz-p2rs",
	journal = "Phys. Rev. D",
	volume = "112",
	number = "8",
	pages = "083024",
	year = "2025"
}

@article{Bronzwaer:2020kle,
    author = "Bronzwaer, Thomas and Younsi, Ziri and Davelaar, Jordy and Falcke, Heino",
    title = "{RAPTOR II: Polarized radiative transfer in curved spacetime}",
    eprint = "2007.03045",
    archivePrefix = "arXiv",
    primaryClass = "astro-ph.HE",
    doi = "10.1051/0004-6361/202038573",
    journal = "Astron. Astrophys.",
    volume = "641",
    pages = "A126",
    year = "2020"
}

@article{Younsi:2019iee,
    author = "Younsi, Ziri and Porth, Oliver and Mizuno, Yosuke and Fromm, Christian M. and Olivares, Hector",
    title = "{Modelling the polarised emission from black holes on event horizon-scales}",
    eprint = "1907.09196",
    archivePrefix = "arXiv",
    primaryClass = "astro-ph.HE",
    doi = "10.1017/S1743921318007263",
    journal = "IAU Symp.",
    volume = "342",
    pages = "9--12",
    year = "2020"
}

@article{Pu:2018ute,
    author = "Pu, Hung-Yi and Broderick, Avery E.",
    title = "{Probing the innermost accretion flow geometry of Sgr A* with Event Horizon Telescope}",
    eprint = "1807.01817",
    archivePrefix = "arXiv",
    primaryClass = "astro-ph.HE",
    doi = "10.3847/1538-4357/aad086",
    journal = "Astrophys. J.",
    volume = "863",
    pages = "148",
    year = "2018"
}

@article{Aimar:2023vcs,
    author = "Aimar, N. and Paumard, T. and Vincent, F. H. and Gourgoulhon, E. and Perrin, G.",
    title = "{GYOTO 2.0: a polarized relativistic ray-tracing code}",
    eprint = "2311.18802",
    archivePrefix = "arXiv",
    primaryClass = "astro-ph.HE",
    doi = "10.1088/1361-6382/ad351d",
    journal = "Class. Quant. Grav.",
    volume = "41",
    number = "9",
    pages = "095010",
    year = "2024"
}

@article{Zhou_2026,
	doi = {10.3847/1538-4357/ae5e67},
	url = {https://doi.org/10.3847/1538-4357/ae5e67},
	year = {2026},
	month = {may},
	publisher = {The American Astronomical Society},
	volume = {1002},
	number = {2},
	pages = {152},
	author = {Zhou, Fan and Huang, Jiewei and Li, Yuehang and Zhang, Zhenyu and Hou, Yehui and Guo, Minyong and Chen, Bin},
	title = {Nonthermal Synchrotron Emission and Polarization Signatures during Black Hole Flux Eruptions},
	journal = {The Astrophysical Journal}
}

@article{Johnson:2023ngEHT,
	author = "Johnson, Michael D. and Akiyama, Kazunori and Blackburn, Lindy and others",
	title = "{Key Science Goals for the Next-Generation Event Horizon Telescope}",
	eprint = "2304.11188",
	archivePrefix = "arXiv",
	primaryClass = "astro-ph.HE",
	doi = "10.3390/galaxies11030061",
	journal = "Galaxies",
	volume = "11",
	number = "3",
	pages = "61",
	year = "2023"
}

@article{Harris2020NumPy,
	author = {Harris, Charles R. and Millman, K. Jarrod and van der Walt, St\'{e}fan J. and others},
	title = {Array Programming with {NumPy}},
	journal = {Nature},
	volume = {585},
	pages = {357--362},
	year = {2020},
	doi = {10.1038/s41586-020-2649-2}
}

@article{Hunter2007Matplotlib,
	author = {Hunter, John D.},
	title = {Matplotlib: A 2D Graphics Environment},
	journal = {Computing in Science \& Engineering},
	volume = {9},
	number = {3},
	pages = {90--95},
	year = {2007},
	doi = {10.1109/MCSE.2007.55}
}

@misc{OpenAIChatGPT,
	author = {{OpenAI}},
	title = {Introducing {ChatGPT}},
	year = {2022},
	publisher = {OpenAI},
	note = {Accessed 2026-09-03},
	url = {https://openai.com/index/chatgpt/}
}

\end{document}